\documentclass[useAMS,usenatbib]{mn2e}
\usepackage{amsmath}
\usepackage{amssymb}
\usepackage{mathtools}
\usepackage{natbib}
\usepackage{graphicx}
\usepackage{gensymb}
\DeclareGraphicsExtensions{.pdf,.png,.jpg,.ps}

\newcommand{\vn}{\textbf{v}_{\rm n}}
\newcommand{\vp}{\textbf{v}_{\rm p}}
\newcommand{\vpn}{\textbf{v}_{\rm pn}}
\newcommand{\tMF}{\tau_{\rm{MF}}}
\newcommand{\tEk}{\tau_{\rm{Ek}}}
\newcommand{\tvisc}{\tau_{\rm{visc}}}
\newcommand{\DOmega}{\Delta \Omega}
\newcommand{\vpnphi}{v_{\rm pn}^\phi}
\newcommand{\absvpnphi}{\vert v_{\rm pn}^\phi \vert}
\newcommand{\vpnphiav}{\langle v_{\rm pn}^\phi \rangle_{\theta, \phi}}

\title[Hydrodynamic simulations of pulsar glitch recovery]
{Hydrodynamic simulations of pulsar glitch recovery}
\author[G. Howitt, B. Haskell and A. Melatos]
{G. Howitt$^{1}$\thanks{E-mail: ghowitt@student.unimelb.edu.au (GH)}, B. Haskell$^{1}$ and A. Melatos$^{1}$\\
$^{1}$School of Physics, University of Melbourne, Parkville, Victoria, 3010, Australia}
\begin{document}
\date{Accepted XXXX. Received XXXX}

\pagerange{\pageref{firstpage}--\pageref{lastpage}} \pubyear{2014}

\maketitle

\label{firstpage}

\begin{abstract}
Glitches are sudden jumps in the spin frequency of pulsars believed to originate
in the superfluid interior of neutron stars.
Superfluid flow in a model neutron star is simulated by solving 
the equations of motion of a two-component superfluid consisting of a viscous
proton-electron plasma and an inviscid neutron condensate in a spherical 
Couette geometry.
We examine the response of the model to glitches induced 
in three different ways:
by instantaneous changes of the spin frequency of the inner and outer boundaries,
and by instantaneous recoupling of the fluid components in the bulk.
All simulations are performed with strong and weak mutual friction. 
It is found that the maximum size of a glitch originating in the bulk decreases 
as the mutual friction strengthens.
It is also found that mutual friction determines the fraction of the frequency jump
which is later recovered, a quantity known as the `healing parameter'.
These behaviours may explain some of the diversity in observed glitch recoveries.
\end{abstract}


\section{Introduction}

Radio pulsar glitches are sporadic jumps in the spin frequency of pulsars which 
occur against a background of steady electromagnetic spin down.
During a glitch, the spin frequency increases by as much as one part in 10$^5$
\citep{Manchester2011}, 
over a time-scale unresolvable by radio timing experiments
\citep{Dodson2007}.
Glitches are usually followed by a recovery period, during which some 
(though not necessarily all) of the increase in spin frequency is reversed, 
and the pulsar returns to a state of steady electromagnetic spin down 
\citep{Wong2001, Yu2013}.
The impermanent part of the glitch is often fitted by 
multiple decaying exponentials, which can have characteristic timescales 
ranging from minutes to weeks for a single glitch
\citep{Dodson2002}.
Often, glitches are also followed by a change in the spin-down rate,
which can persist for as long as the time between glitches
\citep{Yu2013}.

Glitches provide a window into the interior of neutron stars and inform 
theoretical models of bulk nuclear matter
\citep{Link1999, VanEysden2010}.
Since the work of 
\citet{Baym1969a} and \citet{Anderson1975},
it has been widely believed that glitches originate in the superfluid
interior of a neutron star; see
\citet{Haskell2015}
for a recent review of glitch models.
The temperature is likely low enough for neutrons to form Cooper
pairs and hence an inviscid Bose-Einstein condensate in certain regions 
of the interior
\citep{Baym1971}.
Direct evidence for superfluidity has come from recent
observations of the young neutron star in the supernova remnant Cassiopeia A,
whose current temperature is too high if its current cooling rate has been 
maintained since birth.
A recent transition to a superfluid state in the interior, leading to enhanced 
cooling from neutrino emission, seems to be implied
\citep{Heinke2010, Shternin2011, Page2011, Elshamouty2013}.

A superfluid in a rotating container forms an array of vortices with quantized 
circulation, whose configuration determines the angular velocity of the 
superfluid as a whole
\citep{Tilley1990}.
As the pulsar spins down due to electromagnetic braking, 
there is a hydrodynamical lift force, called a Magnus
force, which pushes the vortices out of the star.
In the absence of obstructions, the Magnus force keeps the superfluid neutrons
in corotation with the charge-neutral electron-proton fluid, which in turn is kept 
in corotation with the crust by the strong magnetic field.
As the superfluid vortices pass through the crystalline crust,
however, it is energetically favourable for vortex cores to overlap with 
crustal ions, which means that they `pin' to lattice sites in the crust 
and decouple from the smoothly decelerating proton-electron fluid
\citep{Alpar1977, Anderson1982, Seveso2014}.
Vortices also pin to magnetic flux tubes in the core
\citep{Mendell1991, Link2012},
due to the superconducting nature of the protons in that region
\citep{Migdal1959,Page2006}.
When a single vortex unpins, it can knock-on surrounding vortices before it repins,
causing them to unpin as well --- a vortex avalanche
\citep{Cheng1988, Warszawski2013}.
The dynamics of such avalanches are well-studied in terrestrial systems, such as 
sandpiles and forest fires
\citep{Bak1987, Turcotte1999},
and also in astrophysical contexts, such as solar flares [see
\citet{Watkins2015} 
for a review].
Recent studies of the statistical distributions of glitch sizes
and waiting times between glitches find consistency between pulsar glitch data
and the superfluid vortex avalanche model
\citep{Melatos2008, Warszawski2011, Warszawski2013, Melatos2015}.

The dynamics of individual vortices is highly complex and has been studied 
extensively for terrestrial superfluids such as $^4$He and Bose-Einstein condensates
\citep{Donnelly1991, Fetter2009}.
Such systems can be modelled microscopically using the Gross-Pitaevskii equation
(GPE).
While the GPE approach is feasible for small systems, it does not scale well
to neutron stars, which contain $\gtrsim 10^{15}$ vortices.
Moreover, the GPE describes weakly interacting systems
such as Bose-Einstein condensates, rather than strongly interacting fermionic 
systems such as neutron stars.
On the other hand, the average inter-vortex spacing is much smaller than the 
neutron star radius, so a neutron star superfluid can be
described in the continuum, or hydrodynamic, limit.
Each fluid element contains many vortices, so the rapid spatial 
variation of the superfluid velocity field around the vortices is smoothed,
and the superfluid vorticity can be defined as a continous quantity
\citep{Hall1956, Hills1977, Andersson2006a}.

In this paper, we simulate the response of a model neutron star to glitches
hydrodynamically.
We start by describing the equations of motion for a two-component
superfluid (section 2), then describe the system geometry and 
numerical method (section 3).
We prepare the system in a state of differential rotation (section 4), then
investigate the reponse of the system to glitches induced in three different ways
in section 5. 
Firstly, we spin up the `crust' (section 5.3); secondly, we recouple
the two fluid components in the bulk (section 5.4); and finally,
we spin up the core (section 5.5).
We study all three types of glitches in regimes where the coupling between the
two fluid components, called mutual friction, is either strong or weak.
The response of the spin frequency to a glitch is dramatically different for
each of the three glitch types. 
We find that mutual friction plays an important role in those glitches which originate in 
the bulk.
The observed behaviour is discussed critically in section 6.

\section{Equations of motion}

We model the neutron star as a system of two coupled fluids: 
a neutron condensate, labelled $\mathrm{n}$, and a charge-neutral fluid of 
protons and electrons, labelled $\mathrm{p}$. 
For the purpose of studying glitch relaxation this is an adequate description
\citep{Sidery2010, Haskell2012, Haskell2014}, 
as electrons can be considered locked to the protons on length-scales larger 
than the electron screening length and time-scales longer 
than the inverse of the plasma frequency \citep{Mendell1991}.  
We have the usual conservations laws for the number densities $n_\mathrm{x}$, 
\begin{equation}
\partial_t n_\mathrm{x}+\nabla_i (n_\mathrm{x} v_\mathrm{x}^i ) = 0 \, ,
\end{equation}
where $\mathrm{x}=\mathrm{n},\mathrm{p}$ indexes the constituent, 
$i$ labels Cartesian components, and we adopt the Einstein convention of 
summing over repeated indices. 
The two momentum equations can be written as 
\citep{Prix04}
\begin{align}
& (\partial_t+n_\mathrm{x}^j\nabla_j) 
(v_i^\mathrm{x} + \varepsilon_\mathrm{x} v_i^\mathrm{yx})
+\nabla_i(\tilde{\mu}_\mathrm{x}+\Phi) 
+ \varepsilon_\mathrm{x} v_\mathrm{yx}^j\nabla_iv^\mathrm{x}_j \nonumber\\
&= (f_i+\nabla^j D_{ij}^\mathrm{x})/\rho_\mathrm{x} \, ,
\end{align}
where $\varepsilon_\mathrm{x}$ is the entrainment coefficient, 
$\tilde{\mu}_\mathrm{x}=\mu_\mathrm{x}/m_\mathrm{x}$ is the chemical
potential $\mu_\mathrm{x}$ scaled by the mass $m_\mathrm{x}$ of each component,
$v_i^\mathrm{yx}=v_i^\mathrm{y}-v_i^\mathrm{x}$ is the relative flow velocity,
and $\Phi$ is the gravitational potential. 
Viscous terms are encoded in the tensor $D_{ij}^\mathrm{x}$;
they are more numerous than in a single Newtonian fluid, 
given the additional degrees of freedom of a multi-fluid system 
\citep{Andersson2006a, HaskellComer12}. 
The forces $f_i$ on the right-hand side represent various other interactions
between the fluids, such as the Lorentz force or, as we see below, 
the mutual friction force.
 
To simplify the problem we take both fluids to be incompressible 
and consider only the shear viscosity acting on the electron-proton 
`normal' fluid
\citep{HaskellComer12}. 
We also neglect the effect of entrainment ($\epsilon_\mathrm{x}=0$). 
In the crust the latter can actually be quite a poor approximation, 
as entrainment coefficients can be large 
\citep{Chamel2012,Andersson2012, Chamel2013, Newton2015}. 
However, in the outer core of the star it is a good approximation 
\citep{Carter06}.

Even with these simplifications, the problem is challenging numerically.
The flow at every point is characterised by 
four quantities: the velocity fields of the proton-electron fluid, 
$\textbf{v}_\mathrm{p}$, and the superfluid neutrons $\textbf{v}_\mathrm{n}$
and their respective chemical potentials, $\tilde{\mu}_\mathrm{p}$ and 
$\tilde{\mu}_\mathrm{n}$. 
The gravitational potential $\Phi$ is taken to be constant and absorbed 
in the chemical potential terms.
In the isothermal, incompressible, constant density regime, the flow is described 
by the dimensionless equations\footnote{By neglecting the entrainment and all viscous
terms but the shear viscosity of the `normal' fluid, the second-rank tensor equation (2)
is reduced to a vector equation. From here on, all equations are written using 
conventional vector notation.}
\begin{align}
\frac{d \vp}{dt} + (\vp \cdot \nabla) \vp
& = - \nabla \tilde{\mu}_\mathrm{p} + \frac{1}{Re} \nabla^2 \vp
+ \frac{1}{\rho_\mathrm{p}}\textbf{F}  \, ,\\
\frac{d \vn}{dt} + (\vn \cdot \nabla) \vn
& = - \nabla \tilde{\mu}_\mathrm{n} - \frac{1}{\rho_n} \textbf{F}  \, , \\
\nabla \cdot \vp & = \nabla \cdot \vn = 0 \, ,
\end{align}
where $\rho_\mathrm{p}$ and $\rho_\mathrm{n}$ are the relative densities of 
protons and neutrons (normalised so that $\rho_\mathrm{p} + \rho_\mathrm{n} = 1$), 
and $Re$ is the Reynolds number.
The mutual friction force, \textbf{F}, is a term that describes the interaction
between the proton and neutron fluids, which arises primarily from the scattering
of electrons off vortex cores 
\citep{Alpar1984}.
It is a combination of the Magnus and drag forces of the form 
\citep{Hall1956}
\begin{equation}
\textbf{F} = \rho_n B \hat{\bomega} \times (\bomega \times \vpn)
+\rho_n B' \bomega \times \vpn \, ,
\end{equation}
where $\bomega = \nabla \times \vn$ is the superfluid vorticity,
$\hat{\bomega}$ is the vorticity unit vector
and $\vpn = \vp - \vn$ is the velocity lag between the two fluids, written as
$v_i^\mathrm{yx}=v_i^\mathrm{y}-v_i^\mathrm{x}$ in equation (2).
The parameters $B$ and $B'$ are related ($B' \approx B^2$) dimensionless constants 
\citep{Andersson2006b}.
Vortex lines have a tendency to resist bending, which results in a tension force
which can be included in (6)
\citep{Hills1977, VanEysden2015}.
For simplicity, we ignore the effect of vortex tension, c.f.
\citet{Peralta2008}.
We also neglect vortex tangles and their effect on (6)
\citep{Gorter1949, Peralta2006b, Andersson2007}.

Equations (3)--(5), although derived to describe a condensate coupled 
to a viscous fluid in a neutron star, 
are formally identical to the Hall-Vinen-Bekarevich-Khalatnikov (HVBK) 
equations generally used to describe a (laboratory) condensate coupled to its 
thermal excitations 
\citep{Hall1956, Chandler1986}, 
thus allowing us to make use of numerical schemes developed for the HVBK formalism
\citep{Henderson1995b, Peralta2008}. 
In this context we note that the superfluid pairing gaps are density dependent 
in a neutron star, so there may be regions in which the superfluid 
transition temperature is small, and thermal excitations are important. 
In the present analysis, given the many other simplifying assumptions, 
we neglect this effect.

\section{Numerical method}

\subsection{Pseudospectral solver}

To solve the equations of motion (3)--(5), a pseudospectral collocation method is used
to discretize the spatial coordinates, and a fractional timestep algorithm is used 
to advance the solution in time
\citep{Canuto1993}.
An explicit algorithm (third-order Adams-Bashforth) is used to solve the nonlinear
terms, while the diffusion terms are solved using an implicit
Crank-Nicolson scheme to maintain numerical stability 
\citep{Boyd2013}.
The spectral solver is the same as that used by 
\citet{Peralta2005, Peralta2008},
who in turn followed
\citet{Bagchi2002}.
The radial coordinate $r$ is expanded as a series of Chebyshev polynomials,
and the angular coordinates $\theta$ and $\phi$ are expanded as Fourier series
with parity correction at the coordinate singularity at the poles.
For a detailed description of the solver, see section 3 and the appendix of
\citet{Peralta2008}.
The computational domain and boundary conditions are described in section
3.2 below.
The equations of motion in this work are slightly different to those in 
\citet{Peralta2005,Peralta2008}, 
as noted in section 2.

\subsection{Initial and boundary conditions}

We solve equations (3)--(5) within a spherical Couette geometry, 
consisting of two concentric spherical boundaries rotating about a common axis.
Figure 1 shows a schematic of the computational domain.
\begin{figure}
\centering		
\includegraphics[scale=0.75]{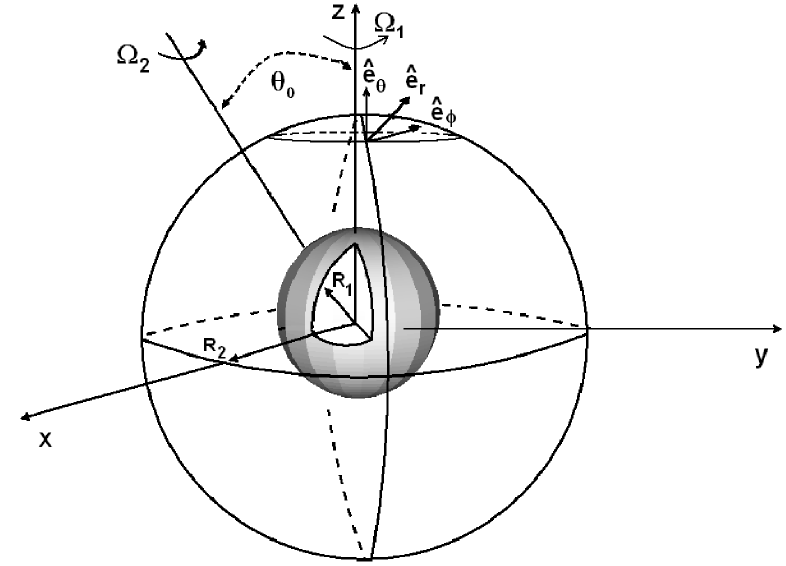}
\caption{Illustration of the model neutron star's geometry.
A spherical surface of radius $R_1$ is embedded inside a larger spherical surface 
of radius $R_2$ with the same origin.
Both spheres may rotate about independent axes, with respective angular 
velocity vectors $\boldsymbol{\Omega}_1$ and $\boldsymbol{\Omega}_2$,
though in this paper we only consider rotation about a common axis. 
The Cartesian coordinates are defined so that the inner sphere rotates 
around the $z$ axis, and the orthonomal vectors 
$\hat{\mathbf{e}}_{r, \theta, \phi}$
form the basis of the spherical coordinate system.
Taken from 
\citet{Peralta2006a}.}
\end{figure}
The inner sphere radius is $R_1$, the outer sphere radius is $R_2$,
and the respective angular speeds are $\Omega_1$ and $\Omega_2$.
In spherical coordinates the computational domain is
\begin{align}
 R_1 \leq & r \leq R_2 \\
 - \pi/2 \leq & \theta \leq \pi/2 \\
 0 \leq & \phi \leq 2 \pi \, .
\end{align} 
The domain is discretised in space according to a Gauss-Lobatto quadrature scheme
[\citet{Canuto1993}; also see Appendix A of \citet{Peralta2008}],
and we typically have $N_r \times N_\theta \times N_\phi = 121 \times 120 \times 4$
discretisation points.
The coarse discretisation in $\phi$ is appropriate, since the common rotation
axis of the crust and core prevents non-axisymmetric flow states, a result which
was verified in 
\citet{Peralta2005, Peralta2008}.

In what follows, we express all variables in dimensionless form.
We normalize times with respect to $\Omega_1(t=0)^{-1}$, so that one rotation 
period at $\Omega_1(t=0)$ corresponds to $2\pi$ time units.
Lengths are normalized with respect to $R_1$,
and velocities with respect to $R_1 \Omega_1(t=0)^{-1}$.
The mass normalization, which affects quantities like the density, moment of inertia,
and torque, is discussed below equation (13) in section 4.1.

A spherical Couette geometry is adopted primarily for numerical reasons,
i.e. to avoid the coordinate singularity at $r=0$ and stabilize the evolution;
flows with $R_2 - R_1 \geq 0.5R_1$ are notoriously unstable
\citep{Benton1974, Yavorskaya1980, Nakabayashi2002}.
The geometry is justified physically as an idealized model of either the outer core
region of a neutron star, in which case the outer boundary is the crust/core interface, 
or the inner crust region, in which case the outer boundary is the neutron drip point.
In either case the inner boundary represents some
phase separatrix, below which the fluids are more tightly coupled by (say) a 
rapid increase in the strength of mutual friction with decreasing radius
\citep{Alpar1984}.
Nonetheless, the presence of a rigid inner boundary is an artificial constraint 
in our model system.
In our simulations we typically set the dimensionless gap width to be
$\delta = (R_2 - R_1)/R_1 = 0.2$.
The region between $R_1$ and $R_2$ is filled with a two-component fluid 
with component densities $\rho_p = \rho_n = 0.5$ and a Reynolds 
number of $Re = 500$ for the proton fluid.
Both of these numbers are artificial, most estimates suggest $\rho_n \approx 0.9$
\citep{Lattimer2004},
and $Re$ may be as high as $10^{11}$
\citep{Mastrano2005,Melatos2007},
but these values are chosen for numerical reasons.

At $t=0$, the inner and outer boundaries are corotating at 
$\Omega(t=0) = 1$.
The simulation runs in a reference frame rotating at 
$\Omega_f = \Omega(t=0)$ for all $t$, which introduces Coriolis terms, 
$2 \boldsymbol{\Omega}(t=0) \times \textbf{v}_{{\rm p,n}}$,
on the right-hand sides of equations (3) and (4).
The initial conditions imply
$\vp = \vn = 0$ everywhere at $t=0$.

For both the proton and neutron velocity fields, we impose no-penetration and
no-slip boundary conditions.
Mathematically, these boundary conditions can be expressed as
\begin{equation}
\textbf{v}_{p,n}(R_{1,2},\theta,\phi) 
= R_{1,2}\boldsymbol{\Omega}_{1,2} \times \mathbf{\hat{r}} \, , 
\end{equation}
where ${\boldsymbol \Omega}_{1,2}$ is the angular velocity of the boundary, 
and $\bf{\hat{r}}$ is the radial unit vector.
The use of these boundary conditions is motivated primarily by numerical 
considerations. 
However, there is some physical justification as well.
At $r = R_2$, no-penetration reflects the inability of either fluid 
to flow through the solid crust. 
No-penetration at $r=R_1$ is more artificial.
Viscous processes as well as magnetic coupling keep the proton fluid 
near both boundaries in corotation, implying no-slip for $\vp$.
The boundary condition for the superfluid neutrons depends on the interaction
between individual vortex lines and the surface.
Following
\citet{Khalatnikov1965},
the relative motion between a vortex line with tangential velocity
$\textbf{v}_L$ and a boundary moving with velocity $\textbf{u}$
can be written as 
\begin{equation}
{\bf v}_L - {\bf u} = 
c_1 \hat{\bomega}_n \times ({\bf n} \times \hat{\bomega}_s)
+ c_2 {\bf n} \times \hat{\bomega}_n \, ,
\end{equation}
where ${\bf n}$ is the unit normal to the boundary.
The coefficients $c_{1,2}$ parametrize the amount of slip.
 The two extremal cases are perfect sliding ($c_1 = c_2 \rightarrow \infty$)
and no-slip ($c_1 = c_2 = 0$). 
One is usually obliged to set $c_1$ and $c_2$ on empirical grounds, even for
well-studied and controlled situations involving liquid helium, let alone under
the uncertain conditions present in a neutron star.
\citet{Peralta2005, Peralta2008}
investigated boundary conditions extensively with this solver and encountered 
numerical difficulties for choices other than no-slip
[c.f. 
\citet{Reisenegger1993} and \citet{VanEysden2015} 
for a discussion of alternatives in analytic calculations].
As mentioned above, we consider the the viscous and inviscid components to be
locked together for $r<R_1$ and postulate that the fluid 
immediately adjacent to this region is also strongly coupled, i.e. 
$c_1 = c_2 = 0$ at $r=R_1$.

\subsection{Model assumptions and limitations}

Necessarily, the above model of involves several simplifying assumptions.
Some of these stem from a lack of theoretical consensus about aspects of the 
physics, while others are made in order to make a difficult numerical 
problem more tractable.
We introduce the assumptions as they arise in previous sections.
Here we summarize them together and discuss why they have been made,
how they affect the applicability of our results to real pulsars, and how
future work might refine the model.
\begin{itemize}

\item \textit{Equations of motion}.
The incompressibility condition (5) is justified, because the sound speed in a 
neutron star is much greater than the flow speeds.
By contrast, the assumption of constant, uniform density does does break down in the
outer core modelled here. 
Work is currently under way to adapt the Navier-Stokes solver on which our two-fluid 
solver is based to work with non-uniform densities 
(K. Poon, private communication, 2015).
The absence of entrainment is valid as long as the simulation volume represents
the outer core, where entrainment is weak 
\citep{Carter06}.
However, the inclusion of an entrainment term in equations (3) and (4) is relatively
straightforward and represents a promising direction for future work.
Vortex tension is another effect we have neglected.
\citet{VanEysden2013} showed that in the crust of a neutron star the effect of vortex 
tension is small and confined to a boundary layer, though more recent work
\citep{VanEysden2015}
suggests that vortex tension adds an oscillatory component to glitch recovery.
As with entrainment, it is relatively straightforward to include vortex tension in the 
solver as a natural next step.
\item \textit{Spherical Couette geometry and boundary conditions}. 
In section 3.2 we explain the reasons for using a spherical Couette geometry
with no-slip boundaries at $R_1$ and $R_2$.
This is an obvious limitation, as it restricts the model to a specific region of 
the star and imposes an artificial boundary condition at the
inner boundary, yet it is unavoidable: the solver in its present form 
(section 3.1) is unstable numerically when applied to a complete sphere.
The appropriate boundary conditions for a superfluid in a rotating 
spherical container remain unclear and depend on the configuration of the 
vortex array and its interaction with the boundary
\citep{Khalatnikov1965, Henderson1995a, Peralta2008}.
Theoretical work by 
\citet{Campbell1982}
suggests that the vortex-boundary interaction is important in a spin-down context,
as the rate of vortex nucleation is greater for rough boundaries than for smooth 
boundaries,
so a rough boundary may decrease spin-down rate of the condensate.
Future laboratory experiments with superfluid helium may shed some light on 
appropriate boundary conditions, but there is no guarantee that the results of 
such an experiment would be applicable to a fermion condensate in a neutron star.
Experiments with liquid helium demonstrate that vorticity is transported erratically
instead of smoothly across a two-phase boundary, such as at $r=R_1$, in response
to interfacial Kelvin-Helmholtz instabilities, an effect which we do not include
\citep{Blaauwgeers2002, Mastrano2005}.
The use of a solid boundary at $R_2$ is also unphysical, because the crust-core interface
is not sharply delineated, and the solid and superfluid components interact non-trivially
through magnetic and elastic coupling.
\citet{Andersson2011} 
developed a hydrodynamic formalism for investigating the crust-core coupling, which may
be useful in the future for determining appropriate boundary conditions in an 
astrophysical context.
Numerical studies of vortex motion are another avenue through which this issue might be investigated in the future
\citep{Schwarz1985, Adachi2010, Baggaley2012}.
\item \textit{Astrophysical parameters}.
Certain parameters of the model are poorly constrained, because bulk nuclear matter
in the low-temperature, high-density regime of neutron stars cannot be studied easily in
laboratory experiments.
For example, the relative moments of inertia of the crust and the superfluid
depend on the precise equation of state for bulk nuclear matter,
and inferred values from measurements of pulsar glitches only weakly constrain these
(sections 4.1 and 6.2). 
Where possible we use the best current estimates and cite relevant literature,
e.g. the values of the mutual friction parameters $B$ and $B'$ (section 2).
In other places, we use artificial values for numerical stability or for illustrative
purposes, all of which are explained in the relevant sections, e.g. the value of 
$Re$ (section 4.1).
Future work will explore the parameter space more widely, in preparation for new
observational studies of pulsar glitches and gravitational waves
\citep{Weber2007, VanEysden2010}.
\end{itemize}

\section{Steady pre-glitch spin down}

Before investigating the post-glitch response in section 5, we discuss how to
set up the simulation to achieve a realistic initial state leading up to the glitch.

\subsection{Set up}

Many models of glitches posit that, as a neutron star spins down, 
the deceleration of the condensate lags that of the viscous component and the crust, 
due to vortex pinning.
As the lag builds up, the pinned vortices act as a reservoir of angular momentum,
which can recouple with the viscous fluid and crust spasmodically, causing
glitches
\citep{Anderson1975, Warszawski2011, Haskell2015}.
To simulate glitches in this paradigm, we need to prepare the system in a state 
where there is a lag between the neutron and proton velocities prior to a glitch.
There is no analytic solution for high-Reynolds-number spherical Couette 
flow in such a state.
Instead, we begin with the inner and outer boundaries corotating and 
spin down the outer sphere by imposing an external torque, 
analagous to the star's electromagnetic braking torque
\citep{Melatos1997}.
As the external torque spins down the outer boundary, 
the neutron and proton fluids near the no-slip boundary also spin down,
and the deceleration is communicated to the interior by Ekman pumping 
\citep{Melatos2012, VanEysden2014, VanEysden2015}.

In response to the external torque, opposing viscous torques $N_1$ and $N_2$ are  
induced by the gradients of $\mathbf{v}_p$ at $r = R_1$ and $R_2$
\citep{Landau1959}.
At each time step, the viscous torque is calculated by integrating the 
non-vanishing terms of the stress tensor over the surfaces $r = R_1, \, R_2$.
Assuming axisymmetric flow one finds
\begin{equation}
N_{1,2} = \frac{2 \pi R_{1,2}}{Re} \int d \theta  \sin \theta 
\left( \frac{\partial v_{p}^\phi}{\partial r} 
\left\vert\vphantom{\frac{1}{1}}_{r = R_{1,2}}\right.
 - \frac{v_p^\phi}{R_{1,2}} \right) \, .
\end{equation}
The angular velocities of the boundaries, $\Omega_{1,2}$, evolve according to
\begin{equation}
I_{1,2} \frac{d \Omega_{1,2}}{d t} = N_{1,2} + N_{\rm{ext}} \, ,
\end{equation}
where $I_1$ and $I_2$ are the moments of inertia of the inner core and the crust, 
and $N_{\rm{ext}}$ is the external torque.
Equation (13) is solved at each timestep using a third-order 
Adams-Bashforth algorithm, so as to maintain the same time-accuracy as the 
pseudospectral solver.
The moment of inertia is expressed in dimensionless units,
normalized with respect to $(\rho_p + \rho_n)R_1^5$.
This means that the moment of inertia of the fluid in the interior, notionally
regarded as a rigid body, is $I_3 = 2.5$.
We choose the moment of inertia of the crust, 
$I_2$ so that $I_2/I_3$ = 4, and the moment of inertia of the core, $I_1$, to 
satisfy $I_2/I_3 = 4$ and $I_1/I_2 = 10.$
These numbers give the fractional moment of inertia due to the condensate as 
$1.1\%$, which is similar to the value of this ratio in glitching pulsars
\citep{Link1999, Andersson2012}.
The external torque, $N_{\rm ext}$ is chosen so that a decoupled crust with moment of
inertia $I_2$ decelerates at $\dot{\Omega} = N_{\rm ext} / I_2 = -10^{-3}$.
This is artificially high; most pulsars have $\vert \dot{\Omega} \vert \sim 10^{-15}$
\citep{Manchester2005}.
However, it is necessary so that an appreciable lag builds up between the crust
and other components in the duration of a typical simulation.

Three important time-scales in this system are the viscous time-scale,
$\tvisc = Re/\Omega$, the Ekman time-scale,
$\tEk = \sqrt{Re}/\Omega$, and the mutual friction time-scale,
$\tMF = 1/(2 \Omega B)$.
In a neutron star, we are interested in two regimes: the viscous-dominant
regime, $\tEk < \tvisc < \tMF$,
and the mutual friction-dominant regime,  $\tMF < \tEk < \tvisc$, which we refer
to as weak and strong mutual friction respectively for the remainder 
of this paper.
For the strong mutual friction case, we take $B = 0.1$, and for weak mutual 
friction we have $B = 10^{-4}$, which are typical values for Kelvin wave damping
and electron scattering respectively
\citep{Alpar1984, Andersson2006b, Haskell2012}.
We typically have $Re = 500$, which is significantly lower than realistic 
neutron star values of $Re \sim 10^{11}$
\citep{Mastrano2005,Melatos2007}.
In the interest of computational tractability we choose lower values of $Re$ to avoid
turbulence in the viscous component
\citep{Peralta2005}, 
which would require increased spatial resolution, and to reduce
the important time-scales $\tEk$ and $\tvisc$.

Starting from a state of corotation, we impose a constant torque on the outer 
boundary and evolve the system until the inner and outer boundaries are spinning
down at approximately the same rate, i.e. we have 
$\dot{\Omega}_1 \approx \dot{\Omega}_2$. 
We refer to this state as `spin-down equilibrium' and use it as a starting
point for our simulations of glitches.
Note that $\dot{\Omega}_1 \approx \dot{\Omega}_2$ does not
imply $\dot{\textbf{v}}_{pn} \approx 0$.
The majority of the angular momentum in the model pulsar is in the
heaviest region, $r < R_1$, which couples to the crust and proton fluid on
either the viscous or Ekman timescales
\citep{VanEysden2014}.
To reach spin-down equilibrium, we need $t > \tvisc > \tEk$, however,
it is still possible for the lag between the proton and neutron fluids to 
grow, as the maximum value of $\vpn$ is determined by 
mutual friction, and is not reached until we have $t > \tMF$.
Evolving the system until one has $\dot{\textbf{v}}_{pn} = 0$ may seem a more
natural initial condition for a glitch, but in practice $\tMF$ can be prohibitively 
long for numerical experiments.
Moreover, it is unclear whether neutron stars ever reach such a state in reality,
as the glitch trigger mechanism is unknown, hydrodynamical instabilities could
set in long before the velocity lag reaches equilibrium
\citep{Andersson2004},
and stratification acts to maintain a shear
\citep{Melatos2012}.
Similar issues arise in laboratory experiments with liquid helium
\citep{VanEysden2011}.

\subsection{Output and initial state}

Figure 2 shows the evolution of $\Omega_1(t)$, $\Omega_2(t)$, 
$\dot{\Omega}_1(t)$, and $\dot{\Omega}_2(t)$, for $0 \leq t \leq 500$.
\begin{figure}
\centering
\begin{tabular}{l l}
\includegraphics[scale=0.325]{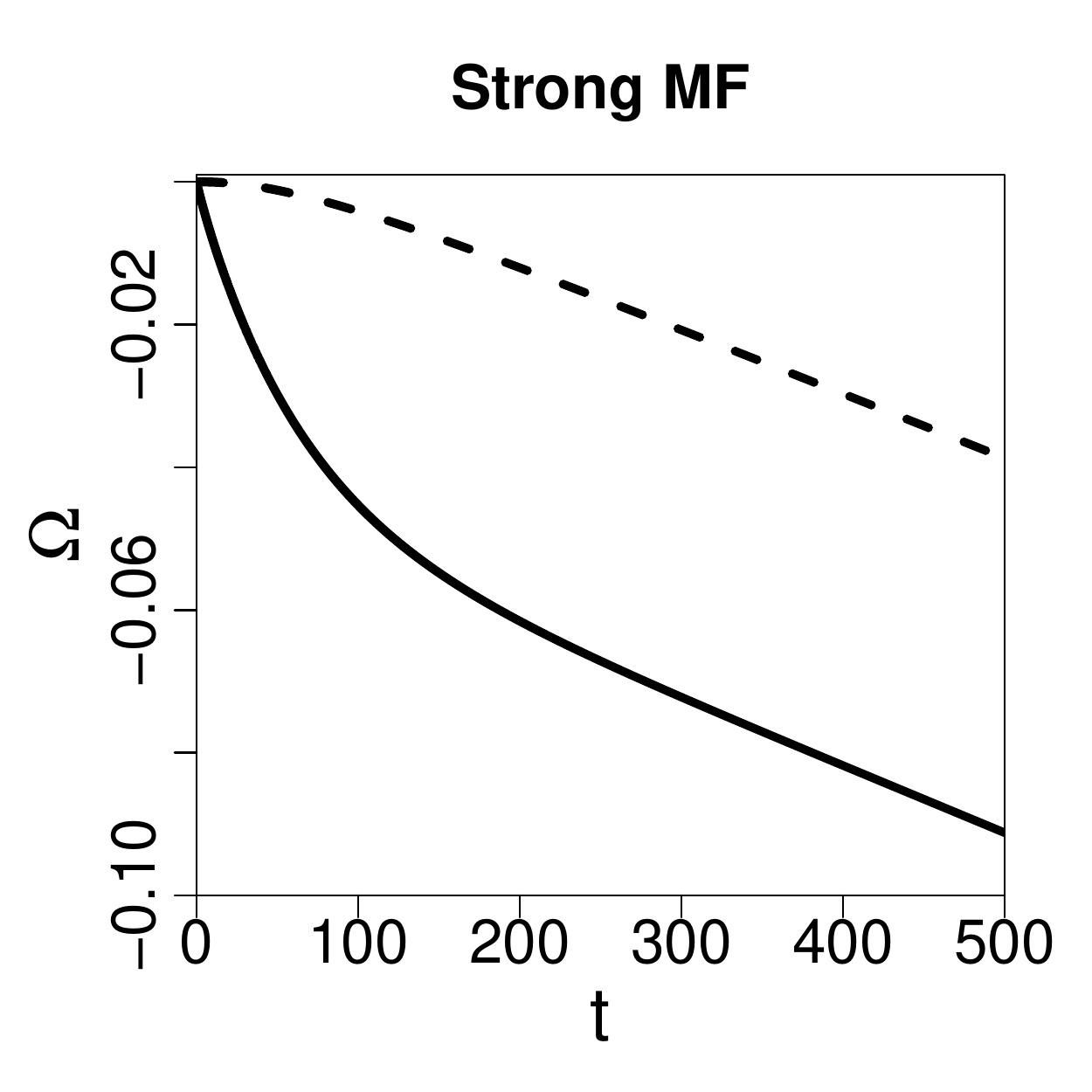} &
\includegraphics[scale=0.325]{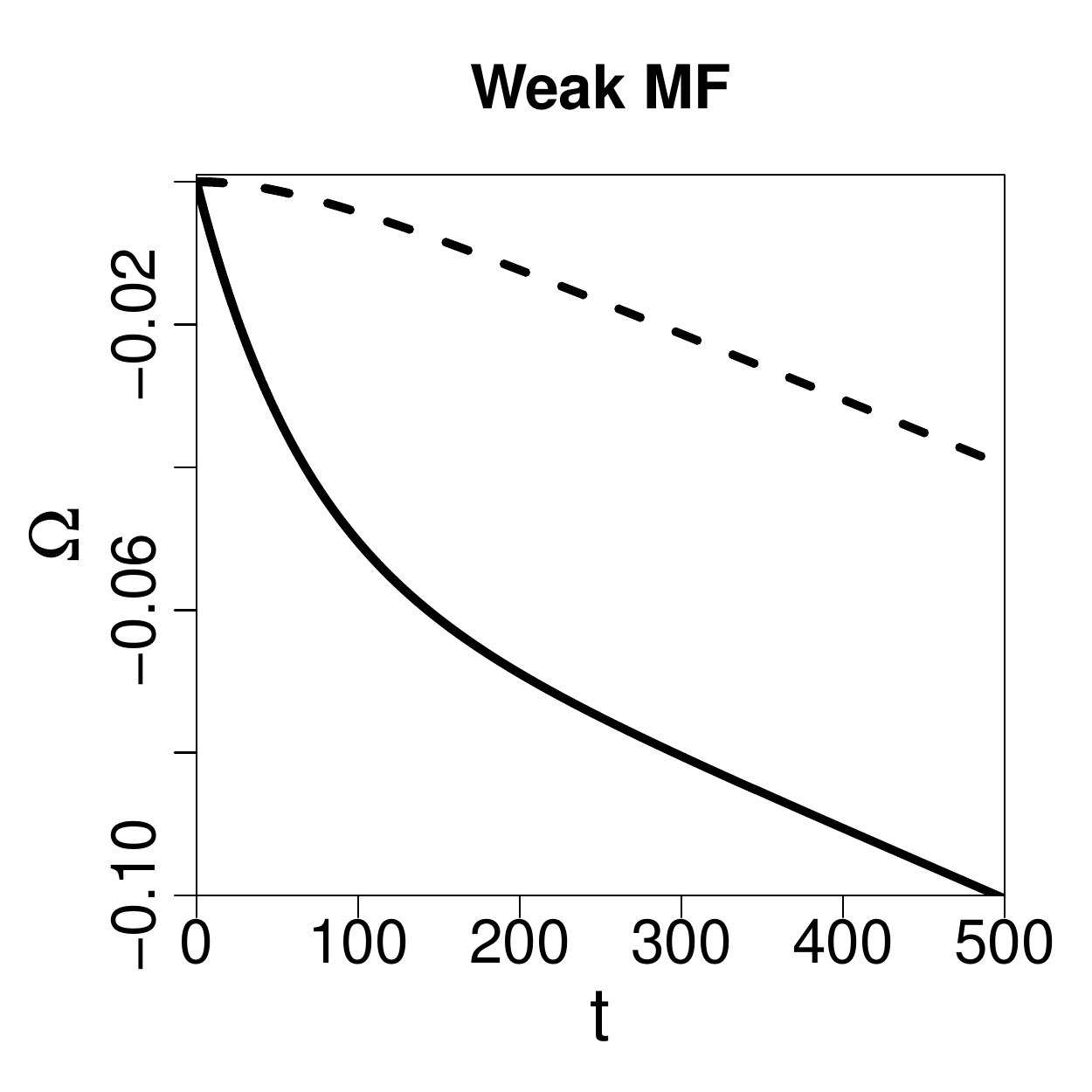} \\
\includegraphics[scale=0.325]{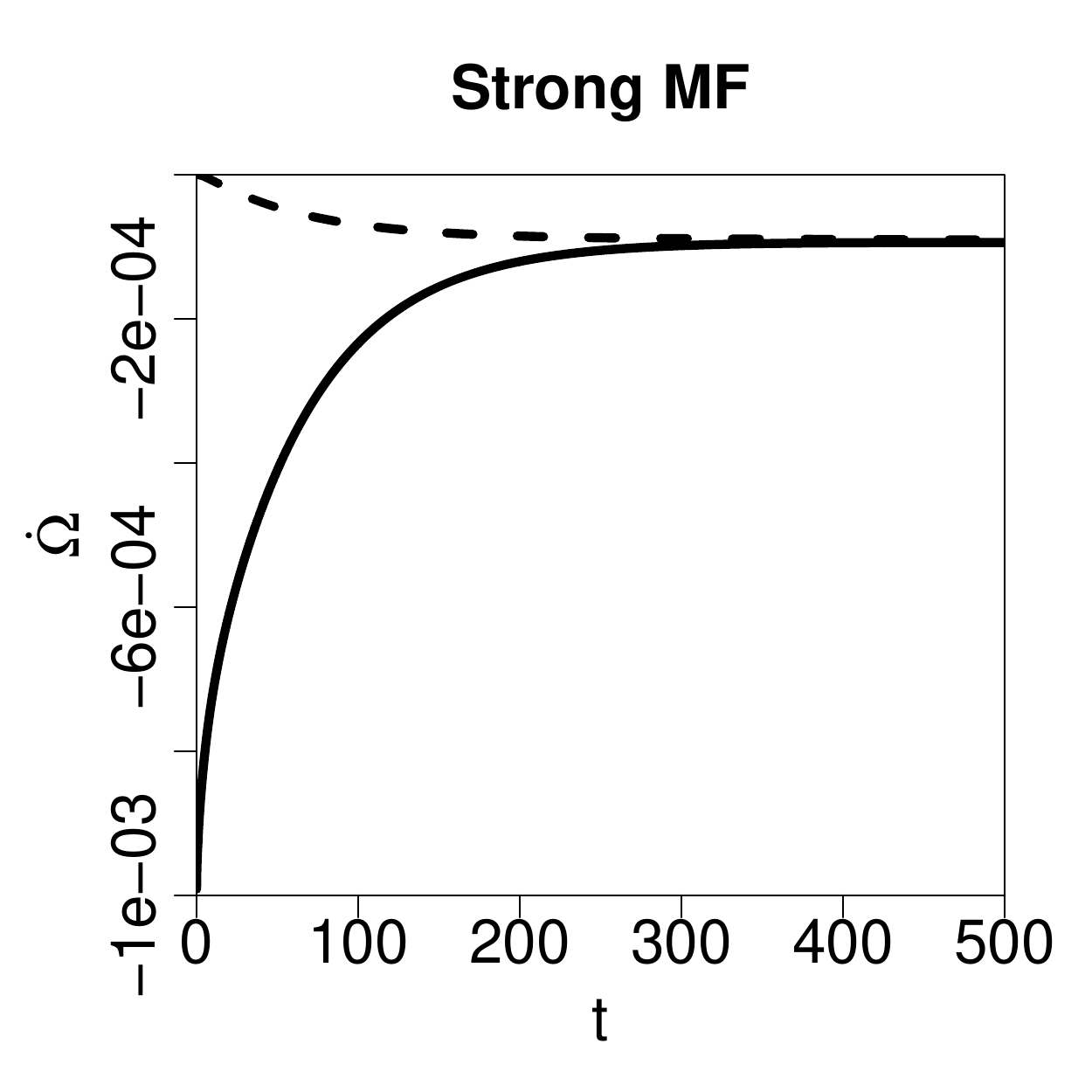} &
\includegraphics[scale=0.325]{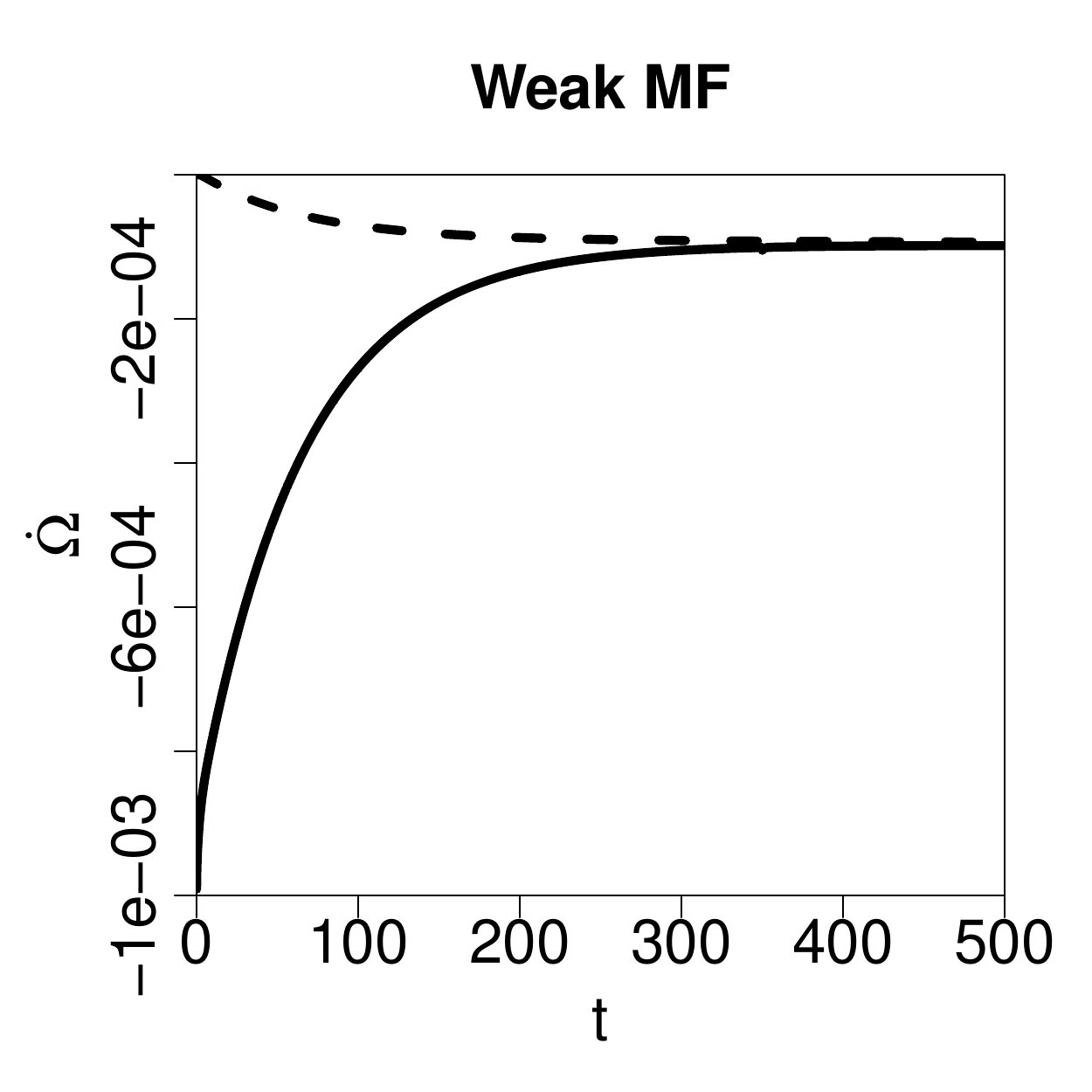}
\end{tabular}
\caption{Response of the crust and core spin frequencies,
$\Omega_1(t)$ and $\Omega_2(t)$, and their time derivatives,
$\dot{\Omega}_1(t)$, and $\dot{\Omega}_2(t)$, to a constant external torque on
the crust, for $0 \leq t \leq 500$, with strong (left column) and weak 
(right column) mutual friction, with $\Omega_1(t=0) = \Omega_2(t=0)$.
In this figure, $\Omega_1$ and $\Omega_2$ are measured in a frame rotating with
angular velocity $\Omega_f = \Omega_1(t=0) = \Omega_2(t=0)$
and are also normalized with respect $\Omega_f$.
Time is normalized with respect to $\Omega_f^{-1}$.
The top row is the crust spin frequency,  $\Omega_1(t)$, (dashed curves), 
and the core spin frequency, $\Omega_2(t)$, (solid curves).
The bottom row is the crust frequency derivative, $\dot{\Omega}_1(t)$, 
(dashed curves), and the core frequency derivative $\dot{\Omega}_2(t)$,
(solid curves).
In both mutual friction regimes the two boundaries are decelerating at
approximately the same rate after $t = 500$, a state we refer to 
as spin-down equilibrium.}
\end{figure}
From the bottom two panels, we can see that the system reaches steady
deceleration by $t = 500$:
we obtain 
$\dot{\Omega}_1 \approx \dot{\Omega}_2 \approx -9.5 \times 10^{-5}$, 
and $\vert \dot{\Omega}_1 - \dot{\Omega}_2 \vert / \vert \dot{\Omega}_1 \vert 
\lesssim 5 \%$, and the angular velocity lag between the crust and core 
$\Omega_2(t = 500)- \Omega_1(t = 500)$ is -0.52
with strong mutual friction and -0.61 with weak mutual friction.

The deceleration is faster with weak mutual friction, as shown in 
the top two panels.
This is because the neutron and proton components couple only after 
$t > \tMF = 500$ (weak), so for $t < 500$ the external torque is only spinning
down the viscous component, the core and the crust.
With strong mutual friction, the inviscid component couples to the viscous
component after $t > \tMF = 5$ (strong), so the moment of inertia 
of the coupled system is greater than with weak mutual friction,
and the magnitude of its deceleration is reduced. 

For a more detailed view of the flow, we look in figure 3 at contours of 
$\vpnphi$ in the strong mutual friction regime in a slice through 
the $y$-$z$ plane, at t = 5 $\approx \tMF < \tEk \ll \tvisc$.
\begin{figure}
\centering
\includegraphics[scale=0.35]{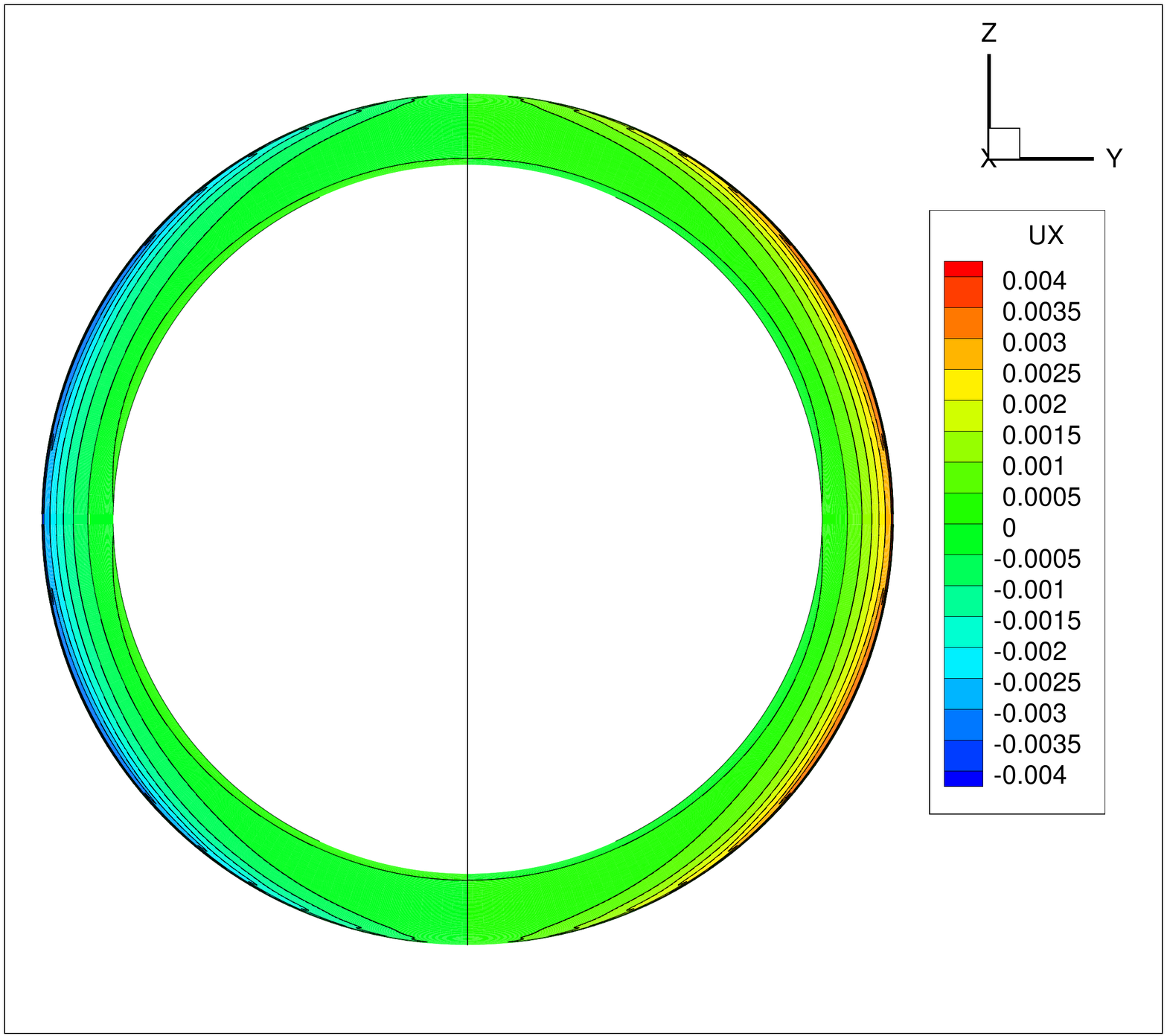}
\caption{Contours of the azimuthal velocity lag, $\vpnphi$,
plotted at $t=5 \approx \tMF < \tEk < \tvisc$ in the $y$-$z$ plane, 
with strong mutual friction.
Velocity is expressed in dimensionless units normalized with respect to 
$R_1 \Omega_f^{-1}$.
$\vpnphi$ is non-zero near $r = R_2$, but decreases monotonically to
zero as $r \rightarrow R_1$.
$\absvpnphi$ increases from zero at the poles to a maximum at 
$\theta \approx 30 \degree$ in latitude.
The flow is axisymmetric but not columnar, as the 18 contours have 
similar curvature to the boundaries throughout most of the domain.}
\end{figure}
It is clear that the flow is axisymmetric but not columnar; 
the contours of $\vpnphi$ are curved.
An expanded view of the flow in the top-right quadrant of the $y$-$z$ plane
appears in Figure 4, at $t = 5$, 25, 100, and 500.
The contours in the remainder of the domain can be 
inferred from the symmetries in Figure 3.
\begin{figure*}
\centering
\begin{tabular}{c c}
\includegraphics[scale=0.25]{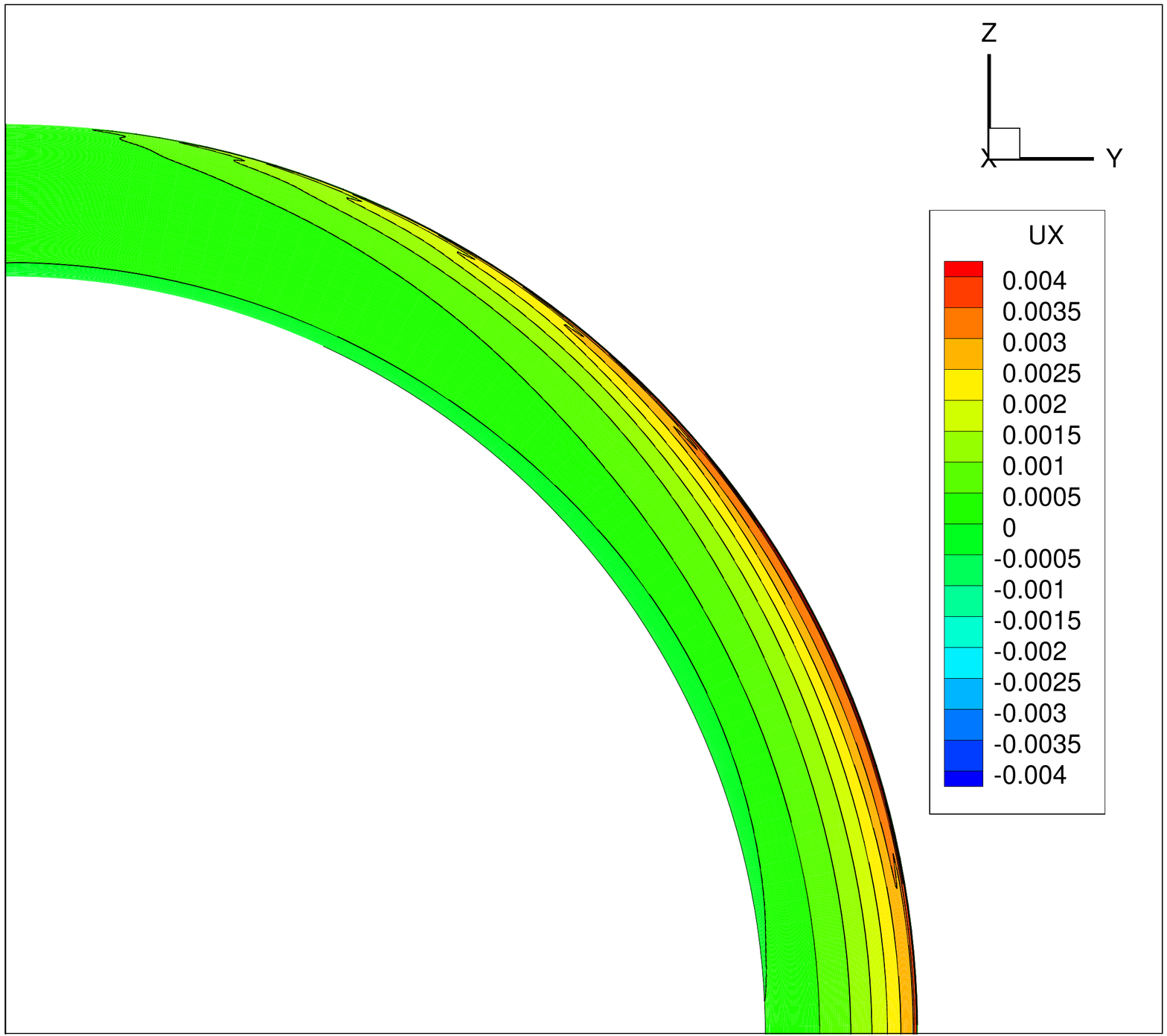} &
\includegraphics[scale=0.25]{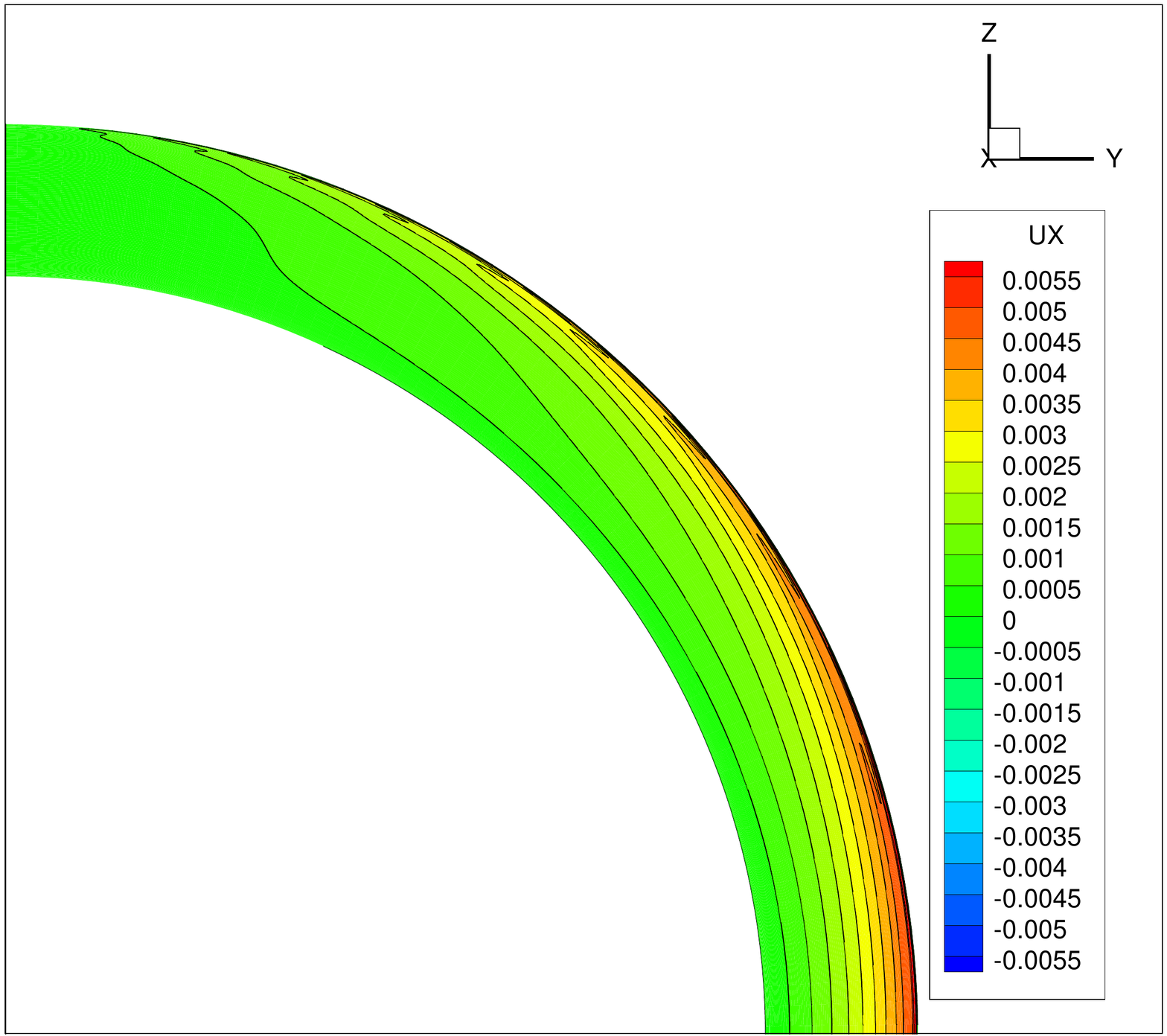} \\
\includegraphics[scale=0.25]{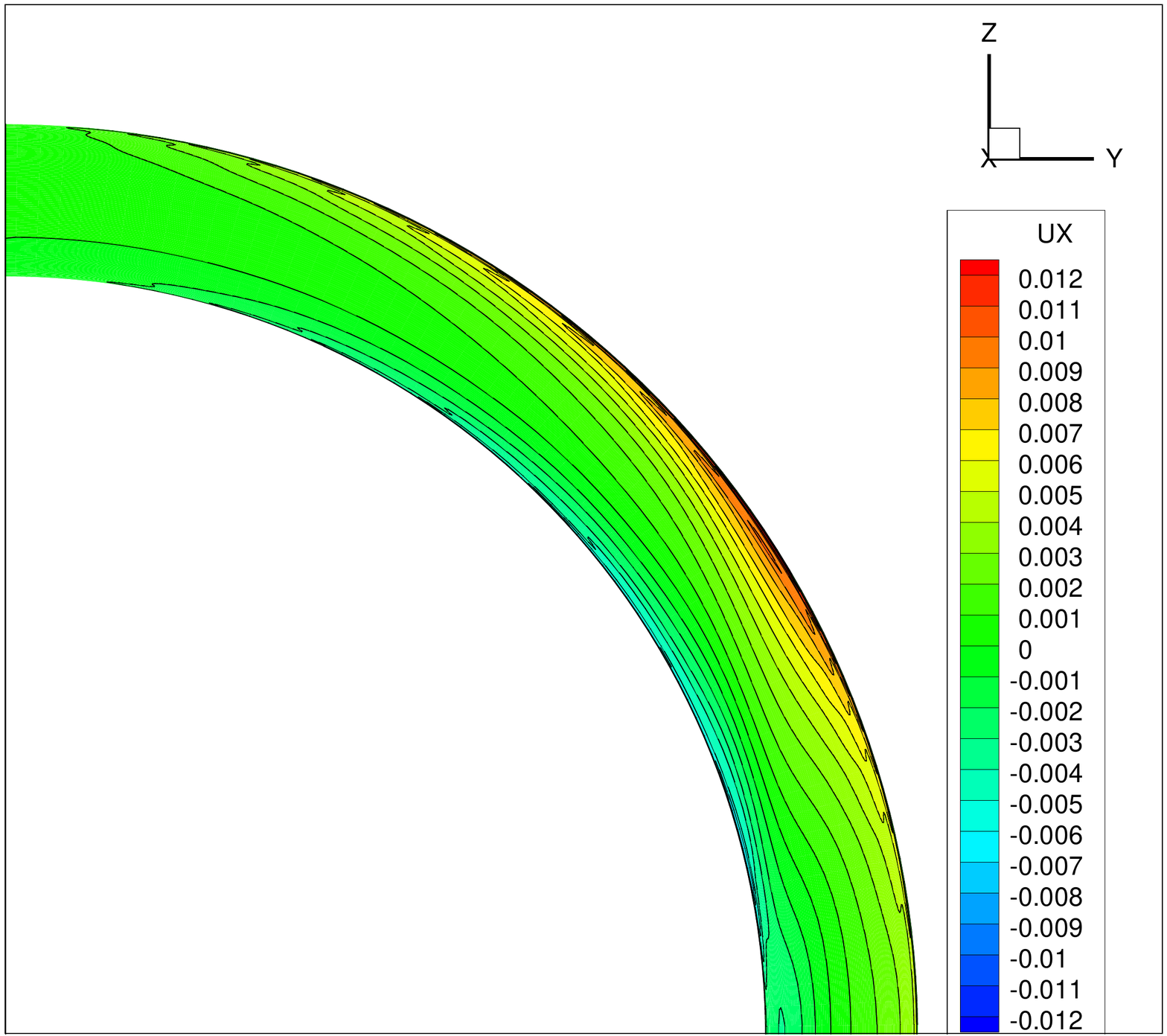} &
\includegraphics[scale=0.25]{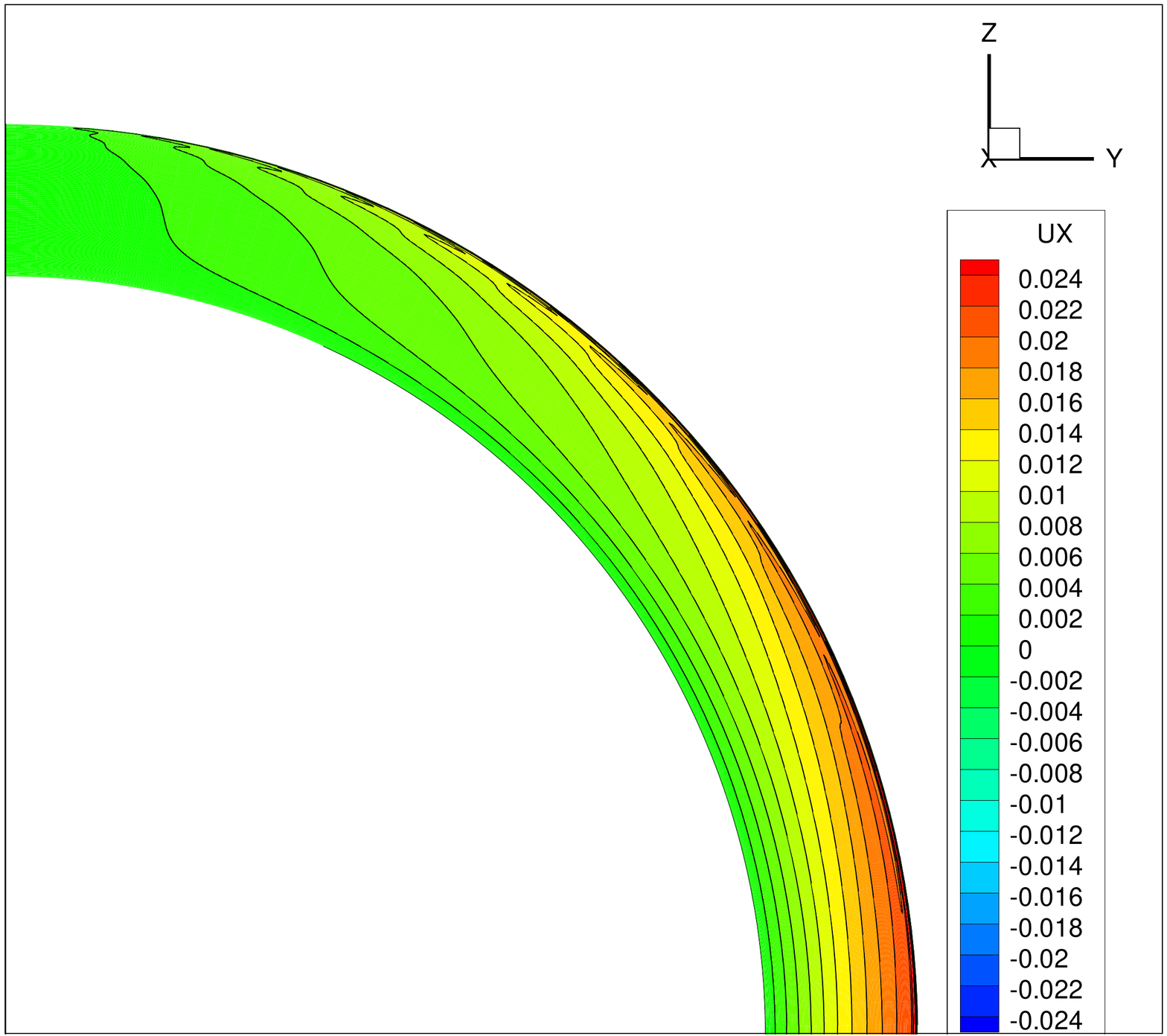} \\
\includegraphics[scale=0.25]{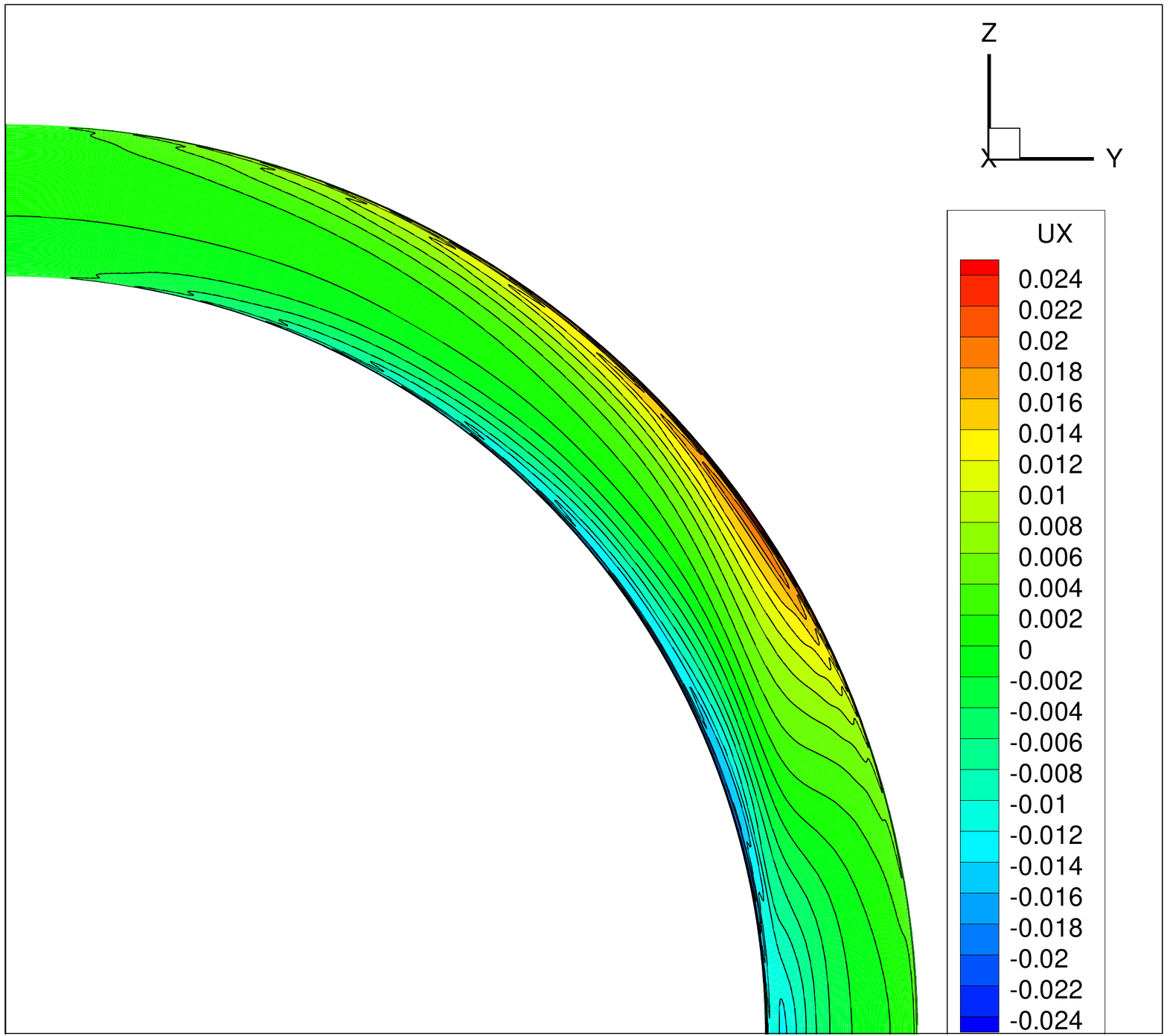} &
\includegraphics[scale=0.25]{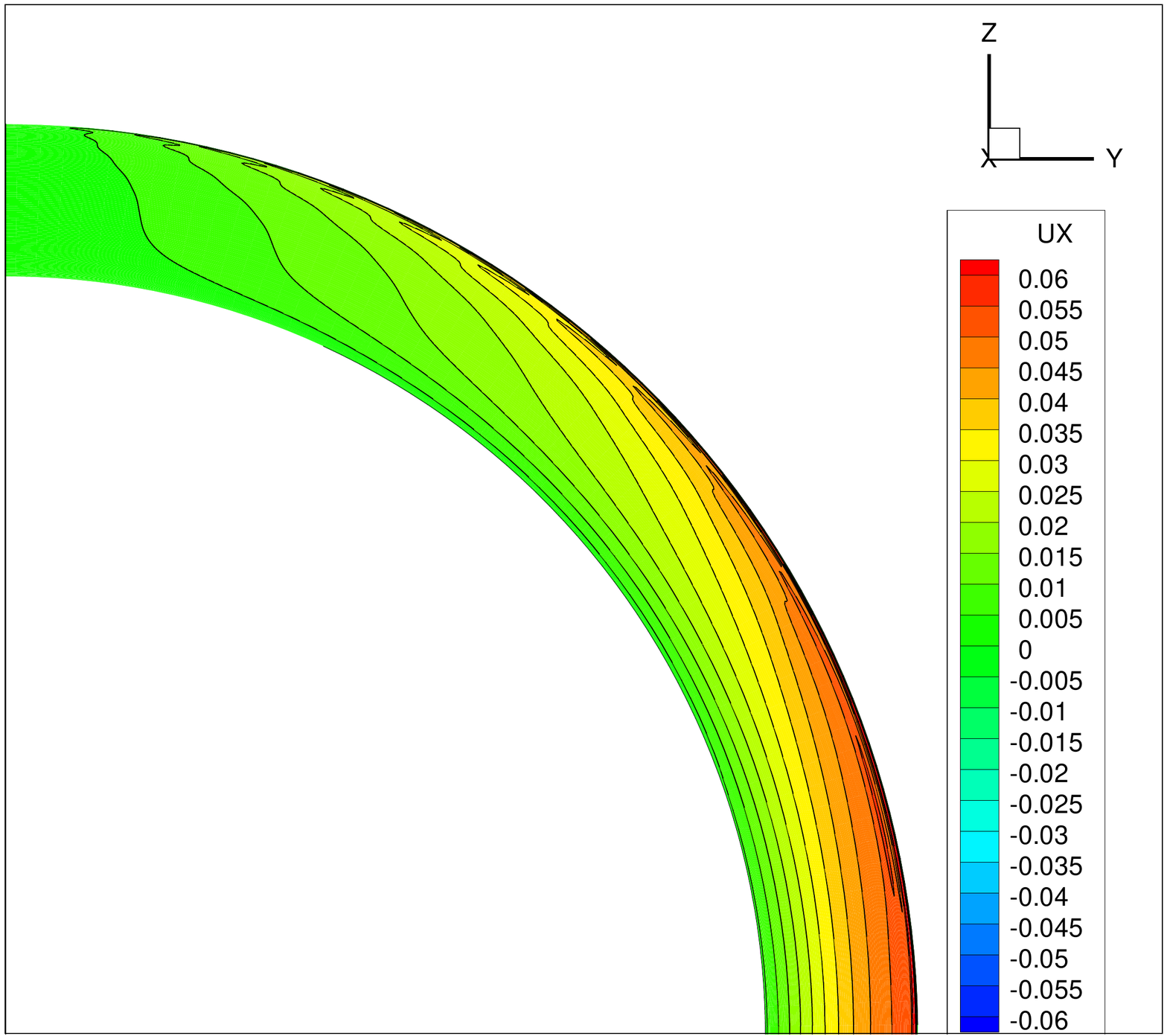} \\
\includegraphics[scale=0.25]{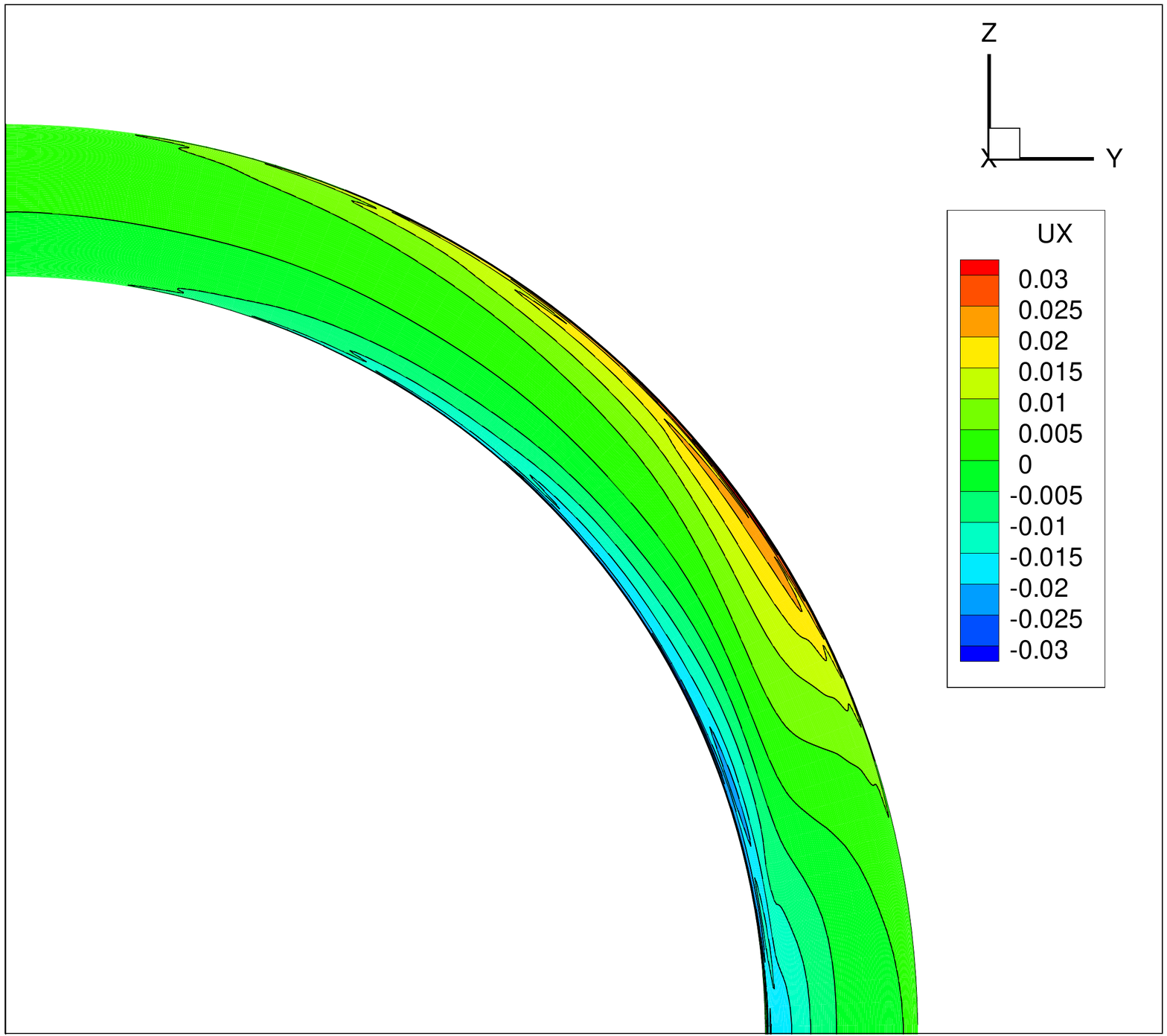}&
\includegraphics[scale=0.25]{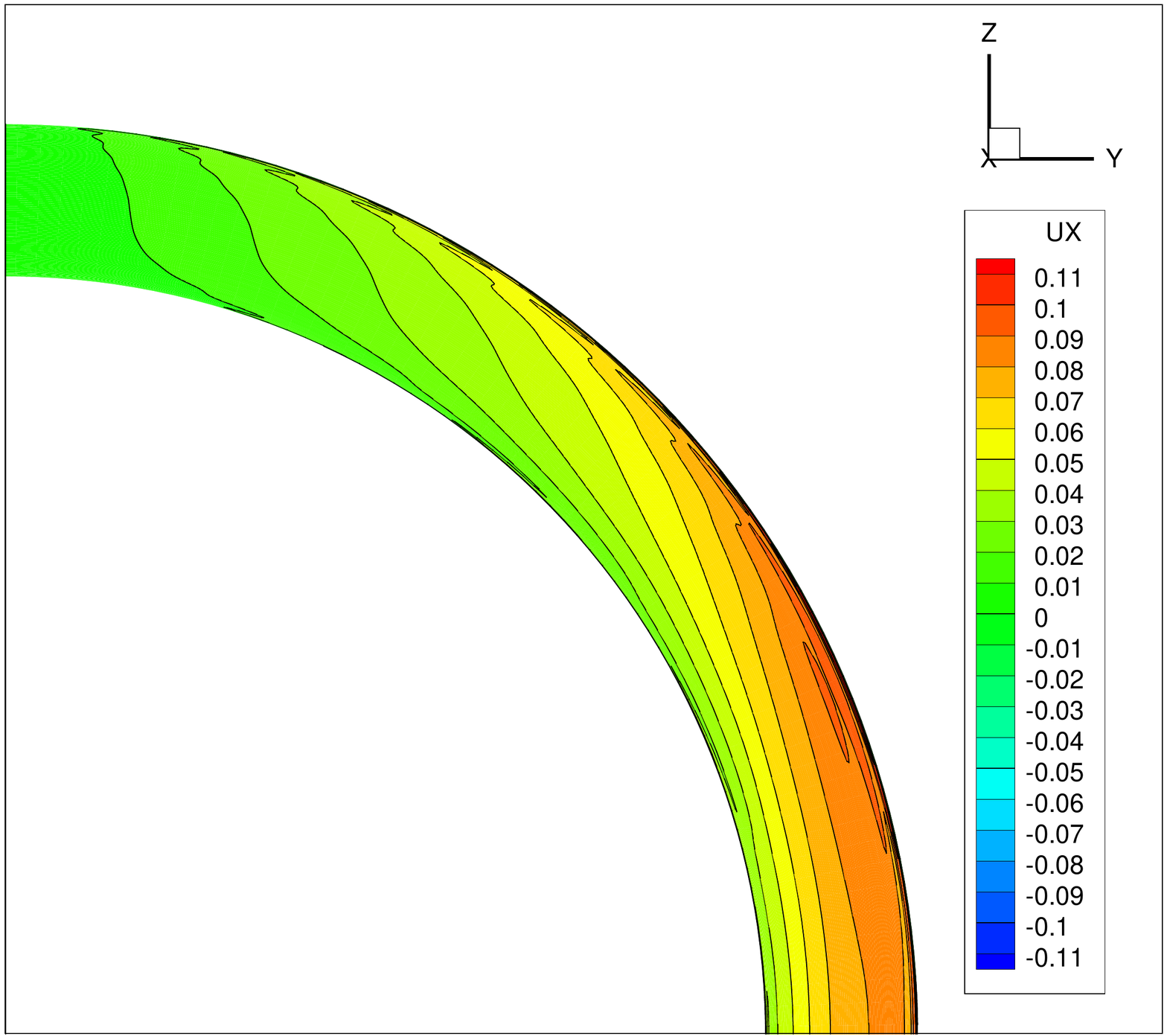} 
\end{tabular}
\caption{Contours of the azimuthal component of the proton-neutron velocity
lag, $\vpnphi$, in the meridional plane, for strong mutual friction
(left column), and weak mutual friction (right column),
at $t = 5$, 25, 100, and 500 (top to bottom row respectively).
Velocity is expressed in dimensionless units normalized with respect to 
$R_1 \Omega_f^{-1}$.
Red colours show positive lag into the page, blue colours show lag out of
the page. 
Rotation is about the positive $z$-axis.
Early on both mutual friction regimes exhibit similar flow patterns: 
The value of $\vpnphi$ is highest near the outer boundary, 
having its maximum value $\theta \approx \pm 30^{\degree}$ in latitude.
Later on, for strong mutual friction, the region where $\vpnphi$ is 
maximised migrates away from the equator, and $\vpnphi$ changes sign
between the inner and outer boundaries.}
\end{figure*}
For strong mutual friction (left column), there is a noticeable difference 
between the flow at $t = 5$ (top row) and at later times. 
At $t = 5$, $\vpnphi$ has a maximum at $r \approx R_2$, at a latitude of
$\theta \approx 30 \degree$.
At fixed radius, $\vpnphi$ decreases slowly with latitude: $\vpnphi(r \approx R_2,
\theta = 0) \approx 0.75 v_{\rm pn, max}^\phi$.
For $t \geq 25$ (second row and below), $v_{\rm pn, max}^\phi$ is still
near $r = R_2$ and $\theta = 30 \degree$, however, $\vpnphi$ falls off more steeply 
with decreasing latitude, with $\vpnphi(r \approx R_2$, $\theta = 0) \approx 
0.25 v_{\rm pn, max}^\phi $ at $t = 25$ and 
$\approx 0.1 v_{\rm pn, max}^\phi $ at $t = 500$.
For weak mutual friction, the flow pattern does not change significantly between
$t = 5$ and $t = 100$, apart from a steady increase in $\absvpnphi$.
Another notable difference between the two regimes is that $\vpnphi$ decreases 
monotonically from  $v_{\rm pn, max}^\phi$ at $r = R_2$ to zero at $r = R_1$ 
for weak mutual friction, while it decreases to $-v_{\rm pn, max}^\phi$ at
$r = R_1$ for strong mutual friction.
To investigate the sign-reversal of $\vpnphi$ more closely, we first average 
$\vpnphi$ over the angular coordinates, $\theta$ and $\phi$, and examine the 
radial profile of the averaged velocity, $\vpnphiav$, in figure 5.
\begin{figure*}
\centering
\includegraphics[scale=1]{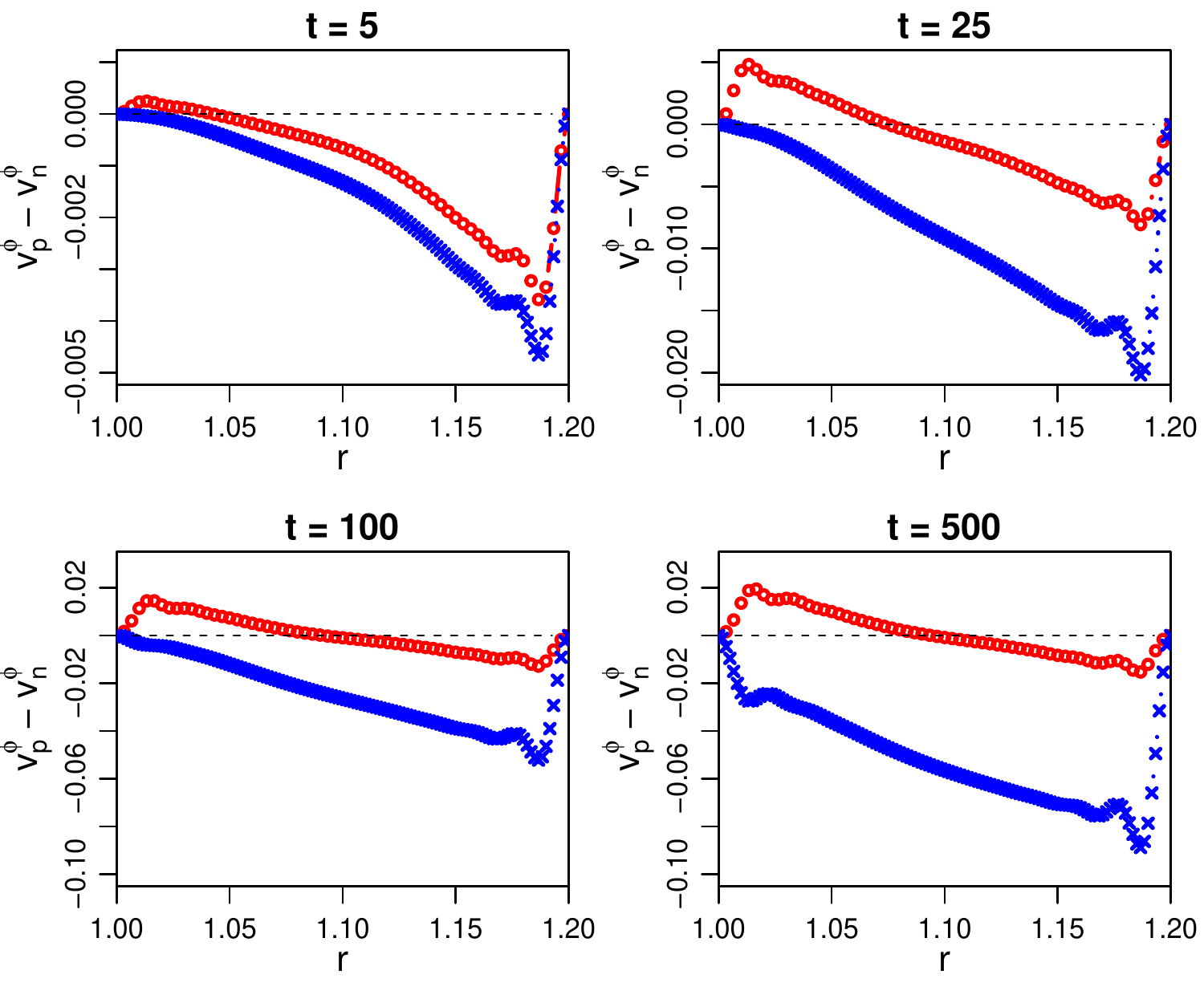}
\caption{Snapshots of the $\theta$ and $\phi$-averaged azimuthal velocity lag,
$\vpnphiav$, versus cylindrical radius $r$, at $t=5$ (top left), 25 (top right), 
100 (bottom left), and 500 (bottom right). 
Time is expressed in dimensionless units normalized with respect to the angular 
velocity of the rotating frame $\Omega_f^{-1}$, 
and velocity is expressed in dimensionless units normalized with respect to 
$R_1 \Omega_f^{-1}$.
The red curve (circles) and the blue curve (crosses) correspond to 
strong and weak mutual friction respectively.
The boundary condition enforces $\vpnphi = 0$ at $r = R_1$ and $r = R_2$.
When mutual friction is strong, $\vpnphi$ changes sign near $r = R_1$.
When mutual friction is weak, the viscous component spins down faster than the
condensate throughout the domain.}   
\end{figure*}
At $r = R_1$ and $R_2$, the two fluids are locked together by the 
no-slip boundary condition.

Away from the boundaries, an interesting feature develops with strong
mutual friction: $\vpnphiav$ undergoes a sign change.
Na\"{\i}vely, one expects the inviscid neutron condensate to lag the protons, 
because Ekman pumping acts on the protons to bring them into corotation with the 
boundaries after $t \approx \tEk$, while there is no equivalent process 
for the condensate.
When mutual friction is weak, this is indeed the case. 
For $\tMF < \tEk$, however, the protons drag the neutrons along as Ekman pumping
proceeds.
Equations (3) and (4) differ in form, so different flow patterns 
develop in the two components 
[\citet{VanEysden2013};
in particular see Figures 1 -- 6 in the latter paper].
In Figure 6, we show the streamlines of the protons (top) and neutrons (bottom)
at $t = 500$ with strong mutual friction. 
It is clear that the flow states are distinct.
\begin{figure}
\centering
\includegraphics[scale=0.325]{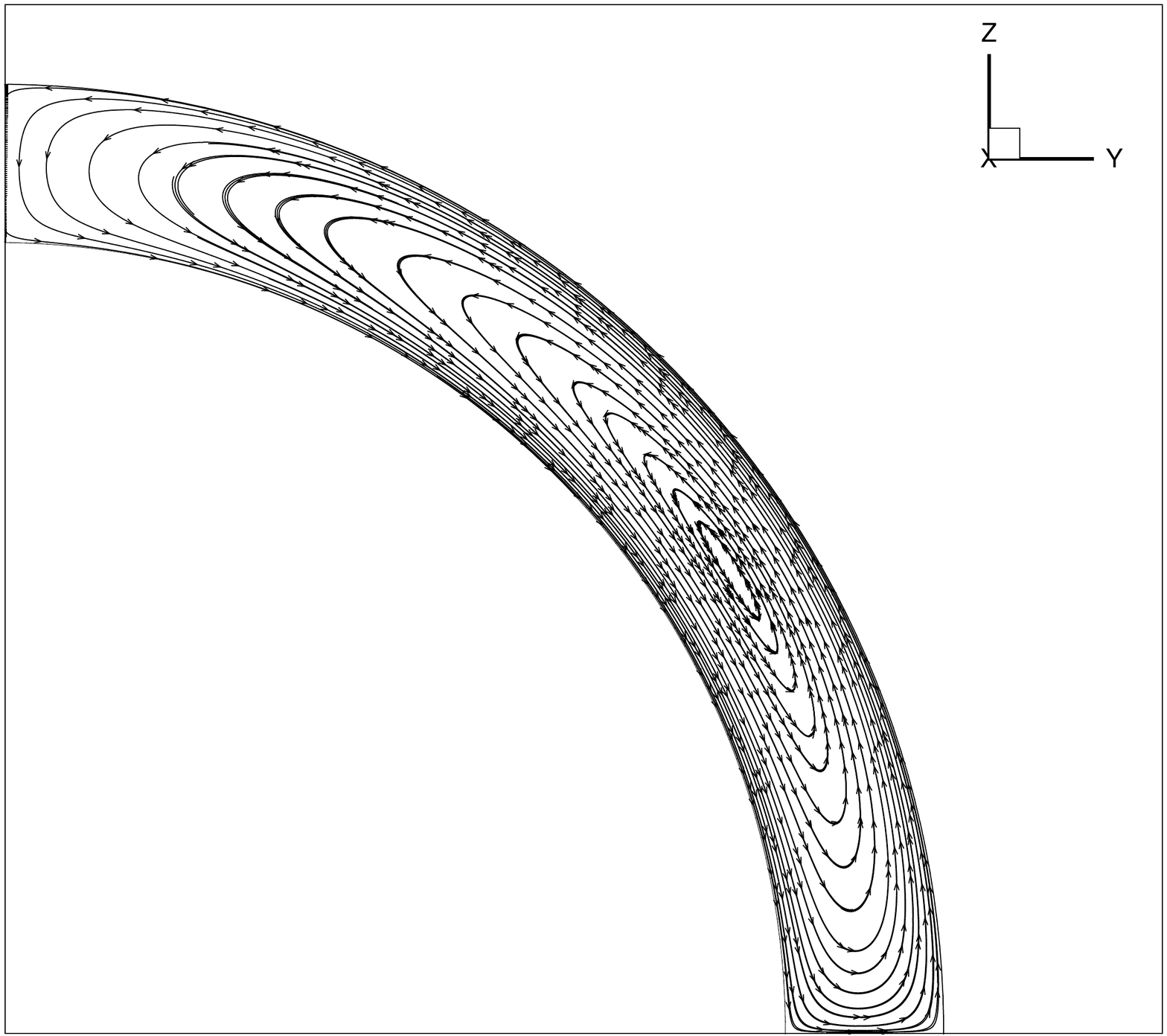} \\
\includegraphics[scale=0.325]{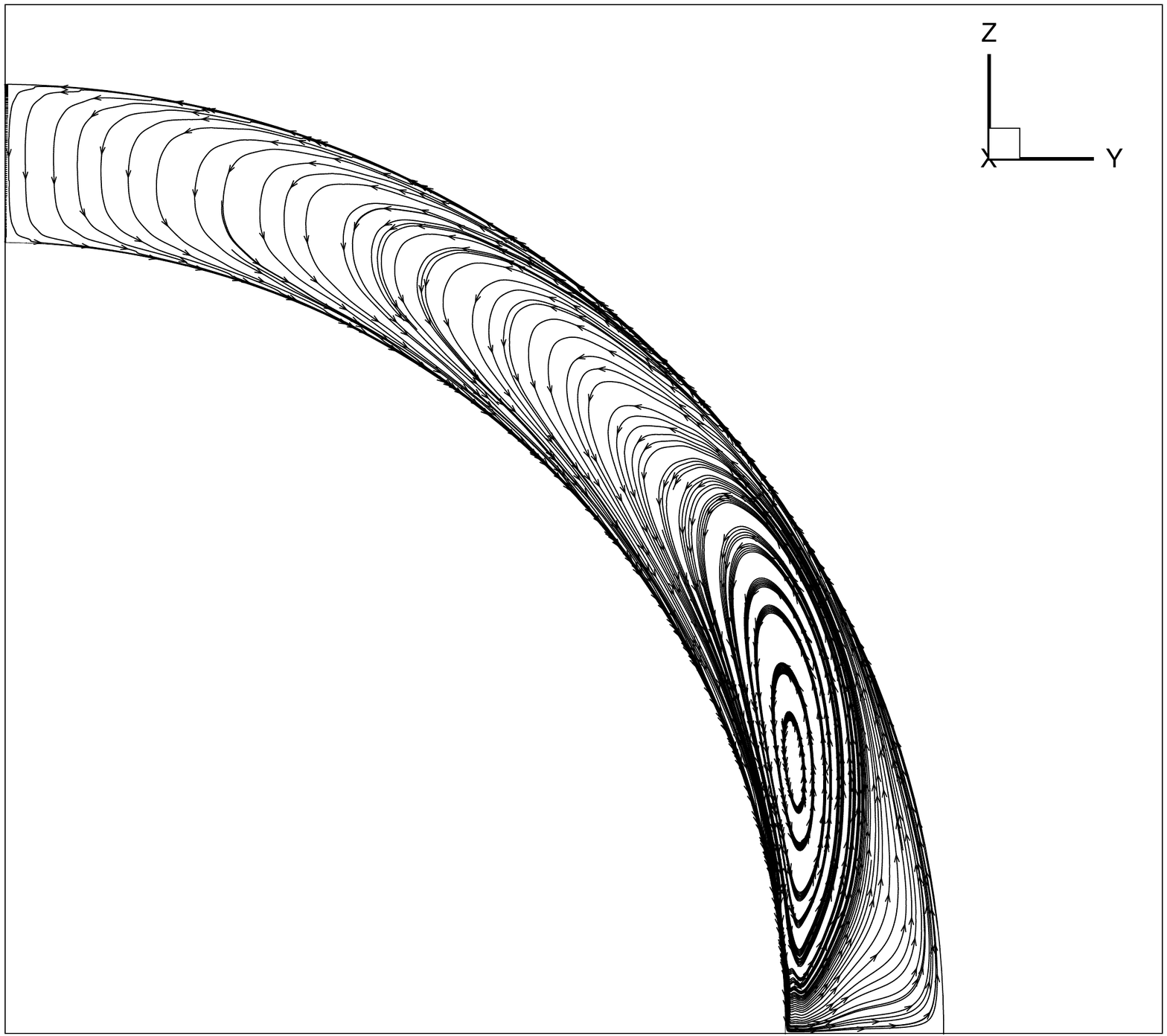}
\caption{In-plane meridional streamlines of the protons (top) 
and neutrons (bottom) with strong mutual friction at $t = 500$.
Both components show meridional circulation, but the flows are distinct.
In the protons the circulation is centered at $\theta \approx 30 \degree$
and $r \approx 0.5(R_1 + R_2)$,
while for the neutrons the circulation cells are centered at 
$\theta \approx 20 \degree$ and $r \approx R_1$.
This difference in flow patterns leads to the result shown in Figure 4,
where $\vpnphi$ changes sign in the region $R_1 < r < R_2$for strong mutual friction. }
\end{figure}
For the proton component, the meridional circulation is centred at 
$\theta \approx 30 \degree$ at $r \approx 0.5(R_1+R_2)$.
The streamlines are parallel to the boundaries at $r \approx R_1$ and $r \approx R_2$
and elsewhere they bend inwards.
For the neutron component, the streamlines are less symmetric.
The primary cell is centered at $\theta \approx 20 \degree$ and is 
closer to $r = R_1$ than $r = R_2$. 
The flow is roughly columnar, with a Stewartson
layer partially forming at $r \sin \theta \approx R_1$
\citep{Peralta2009}.
Subtracting the flow patterns in the two panels of Figure 6 (see bottom left panel
of Figure 4), we find that the neutrons spin down faster than the protons at 
$r \approx R_1$.

\section{Post-glitch recovery}

\subsection{Activating the glitch}

We simulate glitches by taking, as initial conditions, the velocity fields 
and boundary conditions of the system after 500 time units of steady spin down 
and modifying them impulsively in one of three ways.
Firstly, we spin up the outer boundary instantaneously,
which we call a `crust glitch'.
Secondly, we recouple the proton and neutron fluids instantaneously, so that 
$\vpnphi$ is reduced to zero everywhere in a step, which we call a `bulk glitch'.
Thirdly, we spin up the core instantaneously, which we call an 
`inner glitch'.

We can think of a crust glitch as the response to a crustquake, caused by shear stresses
\citep{Ruderman1969, Middleditch2006},
or crust cracking due to a build-up of superfluid vortices at the crust
\citep{Alpar1996}.
Similarly, inner glitches represent a violent event occuring within the core,
at $r < R_1$,
e.g. an avalanche of superfluid vortices pinned to magnetic flux tubes
\citep{Link2012}.
Some models of core matter, such as color superconducting condensates, predict a 
high shear modulus 
\citep{Mannarelli2007}
so that seismic disturbances (`corequakes') may have observable
effects on the crust
\citep{Ruderman1976,Ruderman1998}.

Bulk glitches correspond to trigger mechanisms distributed throughout
the superfluid itself, in the region $R_1 \leq r \leq R_2$.
When $\vert \vpn \vert$ grows too large, a hydrodynamical instability 
similar to the two-stream instability can occur, which acts to reduce the 
velocity lag between the two components
\citep{Andersson2004, Mastrano2005, Glampedakis2009}.
Another potential trigger is a vortex avalanche, driven by the shear between 
the bulk fluid and pinned vortices
\citep{Warszawski2012}.

Following a glitch, we evolve the system for a further 500 time units.
We also run a control simulation, where steady spindown continues uninterupted 
for a further 500 time units, i.e. no glitch.
We compare $\Omega_2(t)$ with and without a glitch and construct `residuals' 
$\DOmega(t)  = \Omega_{2, g}(t) - \Omega_{2, ng}(t)$, 
where  $\Omega_{2, g}(t)$ and $\Omega_{2, ng}(t)$ are the values of $\Omega_2$ 
with and without a glitch respectively.
The residual has two advantages.
First, in reality, glitches are detected as deviations in pulse times-of-arrival 
from a non-glitch spin-down model 
\citep{Hobbs2006, Espinoza2011}.
Second, the residuals separate the glitch evolution from that due to the 
external torque, which remains constant throughout the post-glitch simulation 
(the glitch recovery time-scale is much shorter than the spin-down time-scale).
Note that $\Omega_1(t)$ also evolves, consistent with equation (13).

\subsection{Fitting the glitches}

Generally, pulsar glitches are followed by a recovery phase, during which 
some or all of the increase in spin frequency is reversed.
It is common to fit glitches to a function of the form
\begin{equation}
\Omega(t) = \Omega_0(t) + \DOmega_p  
+ \sum_{i=1}^N \DOmega_n e^{-t / \tau_n} \, ,
\end{equation}
where $\DOmega_p$ is the permanent change in the spin frequency and
$\DOmega_n$ are transient changes in the spin frequency of either sign 
that decay on a characteristic timescale $\tau_n$
[see e.g. 
\citet{Shemar1996, Wong2001}].
Further time derivatives of $\Omega$ may be included in equation (14), 
but are normally neglected.
In section 6.1, we discuss the effect of including a permanent change in the
frequency derivative, $\Delta \dot{\Omega}_p$.
As many as four distinct transient timescales can be fitted for some glitches
\citep{Dodson2007}, 
ranging from tens of seconds to hundreds of days. 
One-dimensional superfuid simulations in 
\citet{Haskell2012}
show that the recovery occurs on a combination of time-scales and is not 
well-approximated by a single exponential, a result which also holds for two-dimensional
Ekman pumping
\citep{VanEysden2010}. 
A multiple exponential fit, though, is well-motivated by theoretical work done by 
\citet{VanEysden2010, VanEysden2014},
who solved analytically for the rotational evolution of a rotating vessel 
filled with a helium-II-like superfluid,
described by the equations of motion of 
\citet{Chandler1986},
following an impulsive acceleration.
They found that, in the limits $\tMF \ll \tEk$, $B' \ll 1$, $I_3/I_2 \gg 1$, 
the recovery involves two exponential timescales, 
which depend on the relative densities of the viscous and inviscid components,
the relative strength of viscous and mutual friction forces, and the 
inertia of the container.

For our simulated glitches, we fit the residual 
$\DOmega(t) = \Omega_{2,g}(t) - \Omega_{2,ng}(t)$ with five paramters:
$\tau_1$, $\tau_2$, $\DOmega_1$, $\DOmega_2$, and $\DOmega_p$.
The fit is extracted using the Levenberg-Marquardt least-squares algorithm, 
implemented in the statistics software \texttt{R} with the package \texttt{minpack.lm}
\citep{Elzhov2013}.

\subsection{Crust glitches}

To induce a crust glitch, we instantaneously increase
$\Omega_2$ by an amount $\vert \Omega_2 - \Omega_1 \vert / 2$ at $t = 500$.
The jump is chosen arbitrarily; we find that its size does not alter 
the dynamics of the recovery qualitatively.
Figure 7 shows the evolution of $\DOmega$ following the crust glitch for strong
(top panel) and weak (bottom panel) mutual friction.
\begin{figure}
\centering
\includegraphics[scale=0.5]{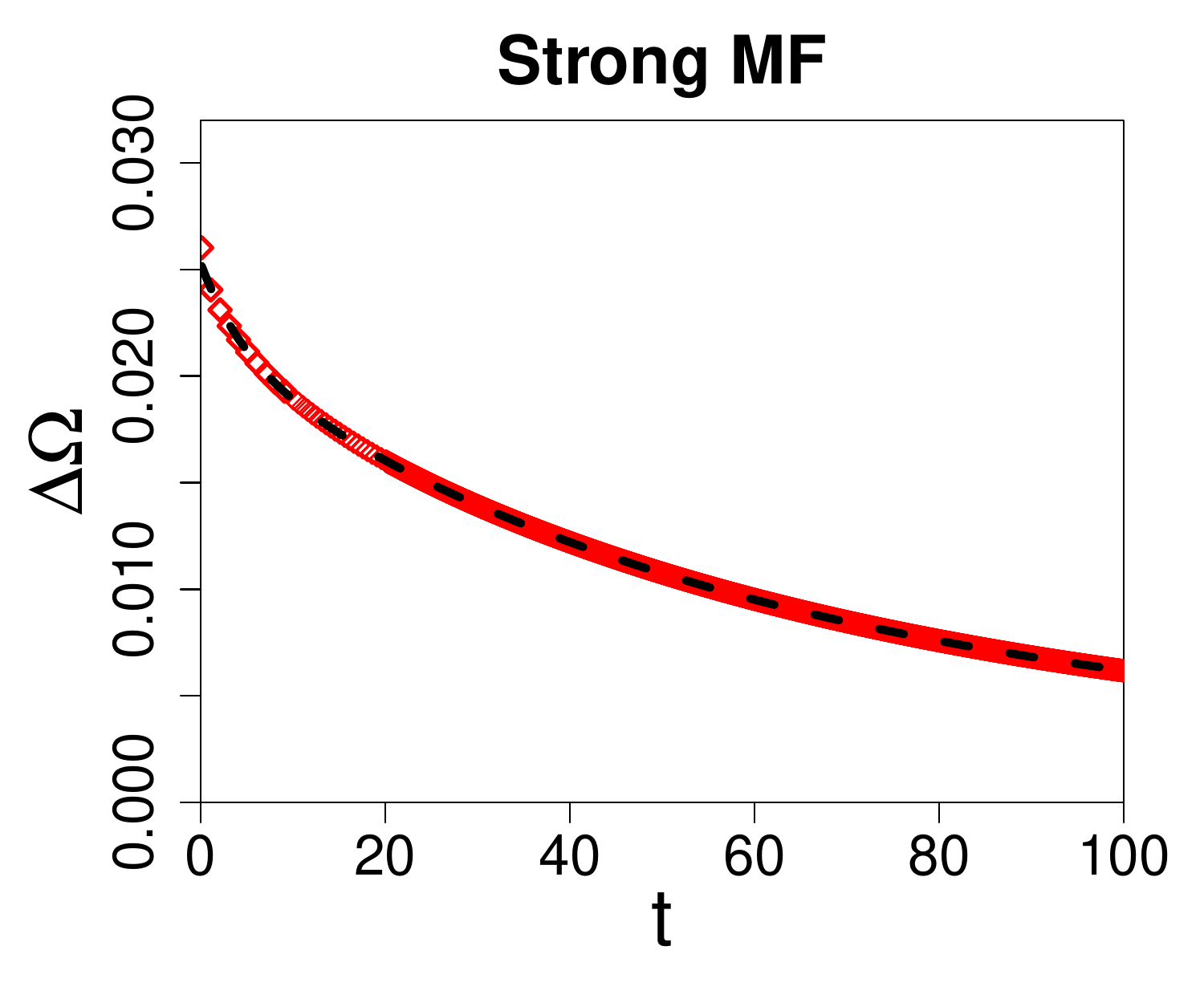} \\
\includegraphics[scale=0.5]{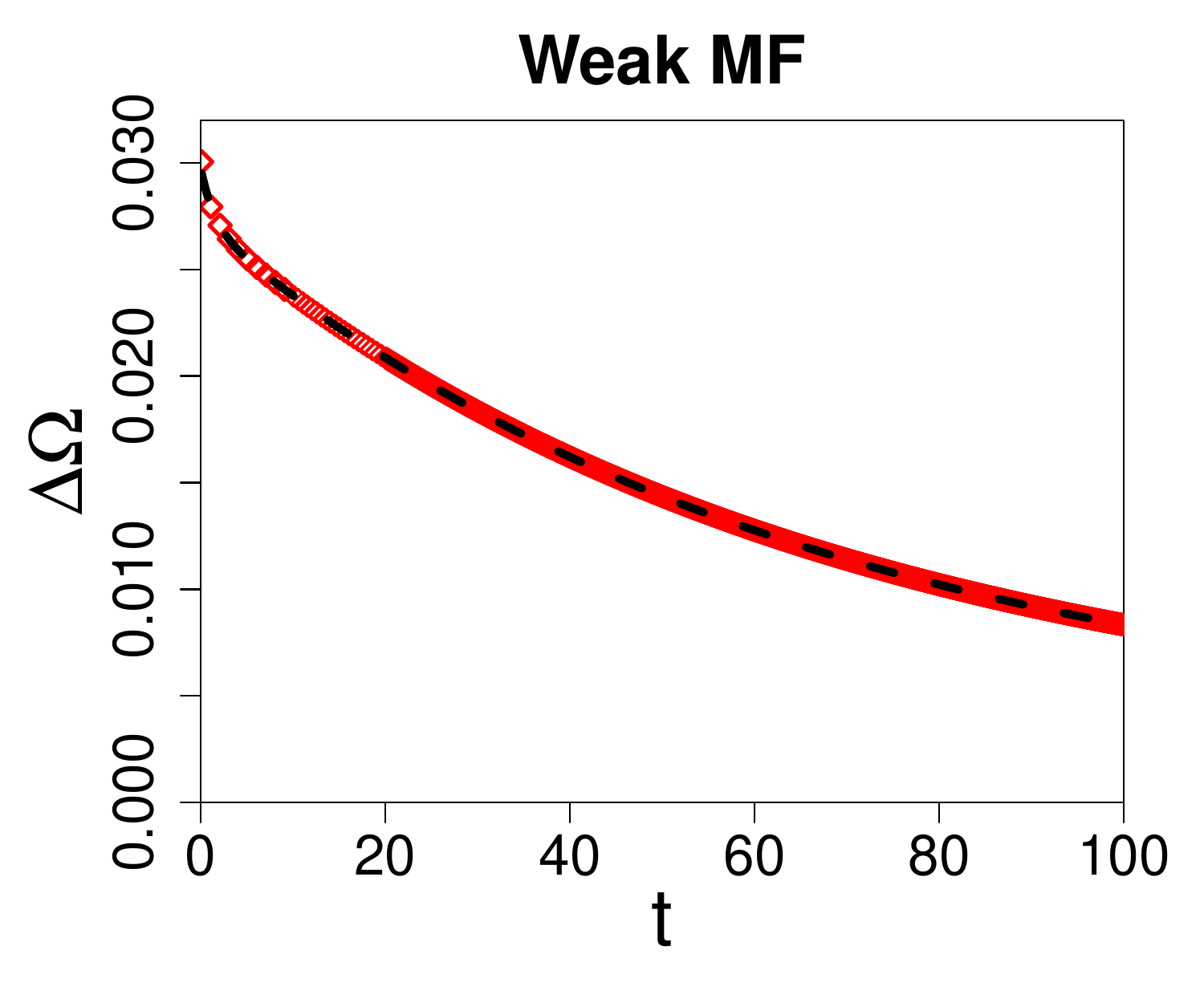}
\caption{Angular velocity residuals,
$\DOmega(t) = \Omega_{2,g}(t) - \Omega_{2,ng}(t)$, for 100 time units following
a crust glitch.
The red diamonds show the data from the simulation.
The black dashed curve shows a curve of best fit to a dual exponential recovery
given by equation (14) with $N = 2$ (see Table 1 for values).
Mutual friction has little effect on the recovery time-scales.}
\end{figure}
The parameters of the associated dual exponential fits are shown in Table 1.
\begin{table}
\centering
\begin{tabular}{l c c }
\hline
Parameter & Weak MF & Strong MF \\
\hline
$\DOmega_1$ & 
$2.55 \times 10^{-3}$ & 
$4.15 \times 10^{-3}$ \\

$\tau_1$ & 
1.53 & 
5.15 \\

$\DOmega_2$ & 
$2.42 \times 10^{-2}$ & 
$1.86 \times 10^{-2}$ \\

$\tau_2$ & 
66.4 & 
62.0 \\

$\DOmega_p$ & 
$2.96 \times 10^{-3}$ & 
$2.46 \times 10^{-3}$ \\
\hline
\end{tabular}
\caption{Fitted parameters to equation (14) with $N = 2$ for the crust glitches 
in Figure 7 with strong ($B = 0.1$) and weak ($B = 10^{-4}$) 
mutual friction (MF) at $Re = 500$.}
\end{table}
Immediately following the glitch $\DOmega$ decreases monotonically.
This happens because the no-slip boundary condition causes $v_p^\phi(r = R_2)$ to also
increase, so the viscous torque, equation (12), is reduced (relative to the no-glitch
simulation), and the external spin-down term in equation (13) dominates.
Both strong and weak mutual friction recover similarly, with $\tau_2 \approx 64$.
The fitted values of $\tau_1$ are discrepant by a factor of $\approx 3$ between
the mutual friction regimes, however, in both cases $\DOmega_1 \ll \DOmega_2$,
so doesn't affect the shape of the recovery profile significantly.
$\DOmega_p$ is $\approx 10\%$ of $\DOmega_{\rm max}$ with strong and weak mutual friction.
There is no qualitative difference between the recovery profile 
in the two mutual friction regimes for crust glitches.
In the limits $\tMF \ll \tEk$, $B' \ll 1$, $I_3/I_2 \gg 1$,
\citet{VanEysden2010} 
derived simple analytic expressions for the time-scales
of a double-exponential recovery, 
\begin{equation}
\tau_1 \approx \tMF \, ,
\end{equation}
\begin{equation}
\tau_2 \approx \frac{\tEk}{\rho_p (1 + I_3/I_2)} \, 
\end{equation}
(or vice versa).
For $B = 0.1$ (strong mutual friction), equations (14) and (15) give 
$\tau_1 = 5$, $\tau_2 = 36$,
compared to $\tau_1 = 5.15$, $\tau_2 = 62.0$ obtained from fitting to simulations.
This is a decent level of agreement, though it should be noted that equations (15) and
(16) are not exactly the expressions as written in 
\citet{VanEysden2010}, 
who considered a spherical container ($R_1 = 0$) filled with a HVBK superfluid
\citep{Chandler1986}, 
rather than a spherical Couette geometry filled with fluid obeying equations (3) -- (5).
More specifically, in \citet{VanEysden2010}, 
the quantity $I_3/I_2$ is the ratio of the moment of inertia of the interior of the
sphere to that of the container, while in our analysis we exclude the moment
of inertia of the core.
We are also in a different parameter regime to
\citet{VanEysden2010}: 
we have $\tEk \approx 5 \tMF$ for strong mutual friction, and $I_3/I_2 \approx 0.25$.

\subsection{Bulk glitches}

To induce a bulk glitch, we instantaneously recouple the proton and neutron fluids,
so that the velocity lag is reduced to zero, in a way that conserves total
angular momentum.
Generally, the recoupling algorithm depends on the relative moments of inertia 
of each fluid, but since we have $I_n = I_p$, we simply set
$v_p^\phi \mapsto v_p^\phi - \vpnphi/2$, and
$v_n^\phi \mapsto v_n^\phi + \vpnphi/2$.
Figure 8 displays the residual $\DOmega$ in the 100 time units of simulation time 
following a bulk glitch with strong and weak mutual friction.
\begin{figure}
\centering
\includegraphics[scale=0.5]{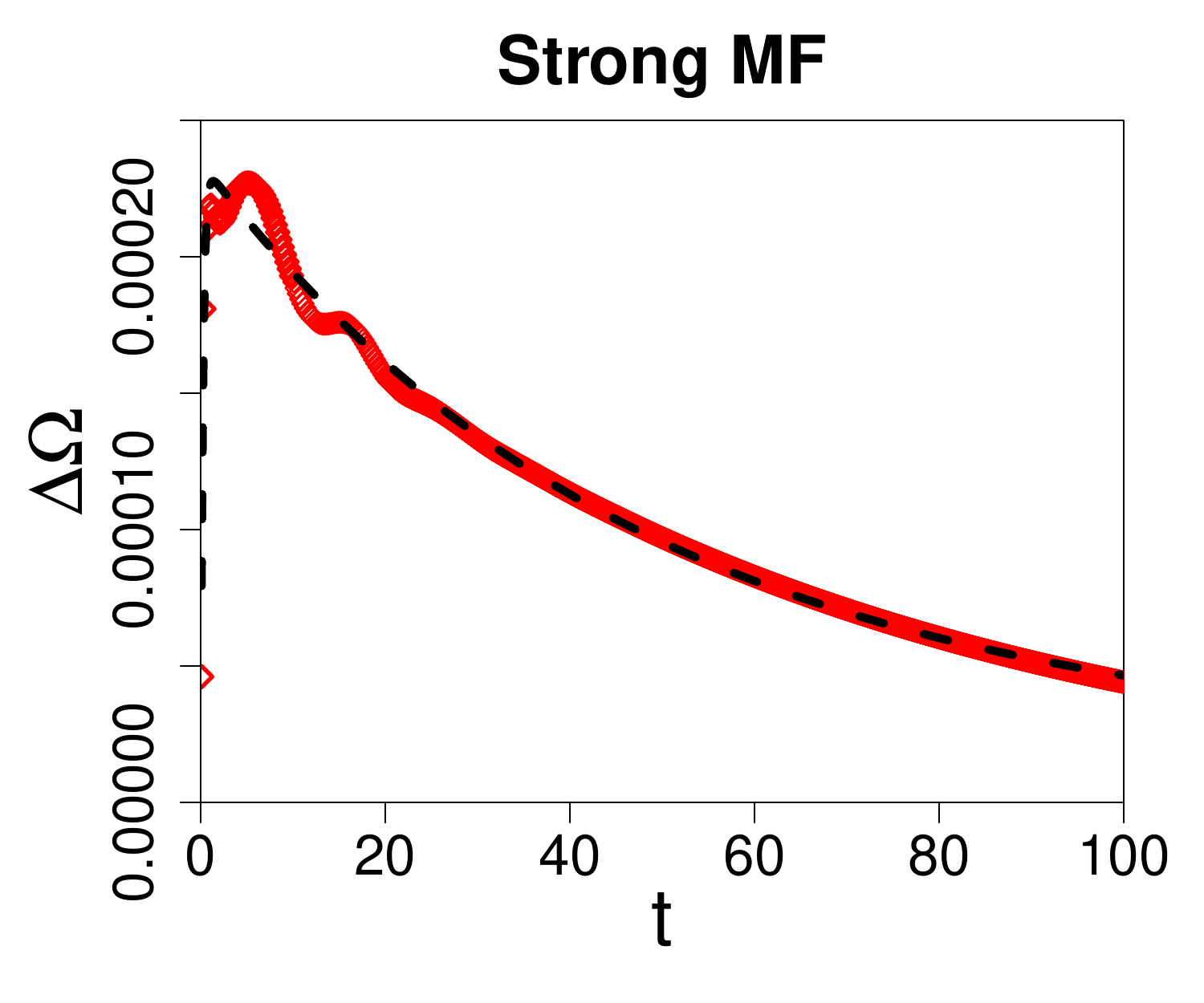} \\
\includegraphics[scale=0.5]{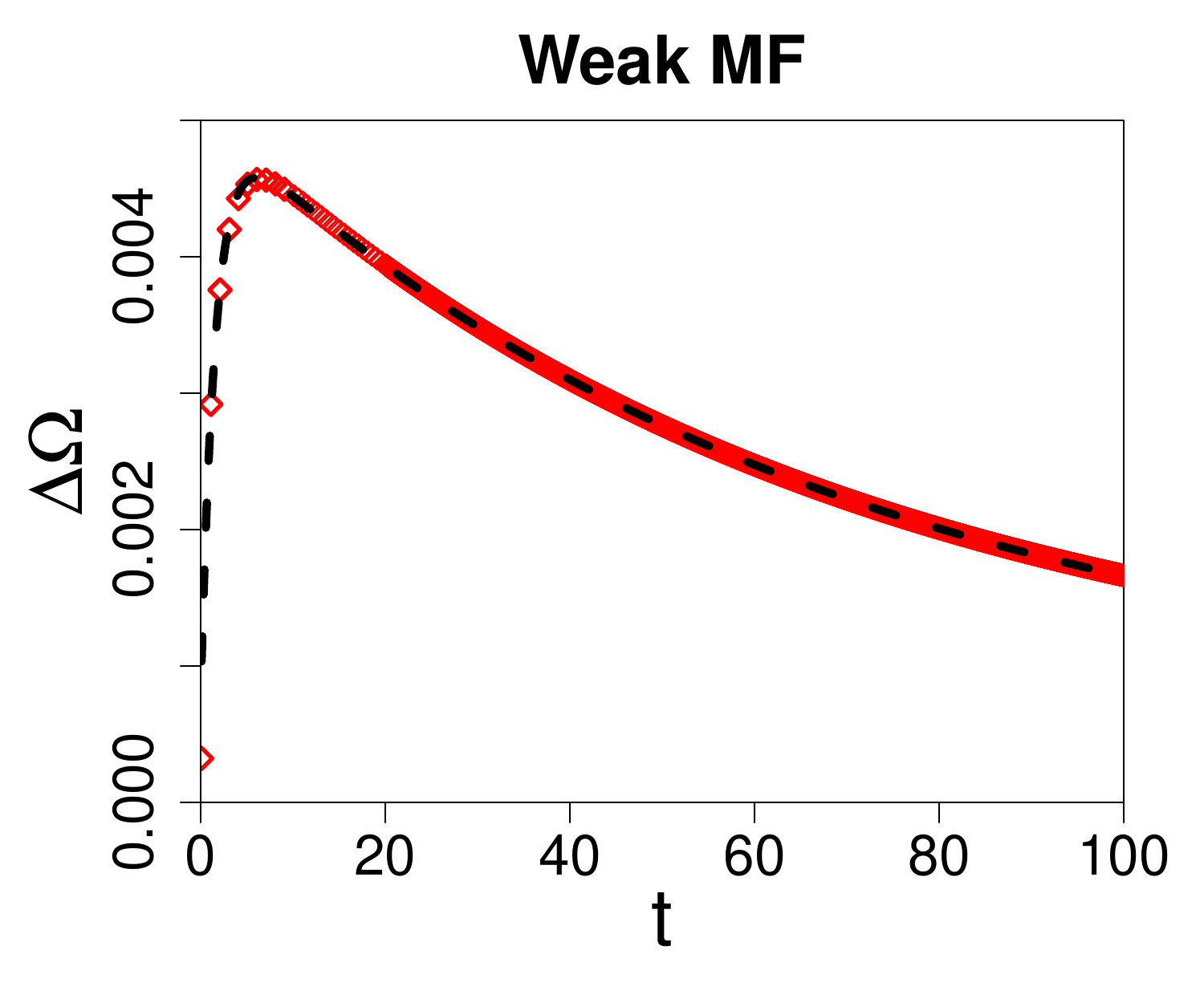}
\caption{Angular velocity residuals,
$\DOmega(t) = \Omega_{2,g}(t) - \Omega_{2,ng}(t)$, for 100 time units following
a bulk glitch.
The red diamonds show the data from the simulation.
The black dashed curve shows a curve of best fit to a dual exponential recovery
given by equation (14) with $N = 2$ (see Table 2 for values).
With strong mutual friction, the size of the glitch is significantly reduced,
and there is an oscillatory component to the recovery for $t \lesssim 30$.}
\end{figure}
The recovery is fitted to equation (14), and the fitted parameters are 
quoted in Table 2.
\begin{table}
\centering
\begin{tabular}{l c c}
\hline
Parameter & Weak MF & Strong MF \\
\hline
$\DOmega_1$  
& $-4.27 \times 10^{-3}$ 
& $-2.61 \time 10^{-4}$   \\

$\tau_1$ 
& 1.60 
& 0.201 \\

$\DOmega_2$  
& $4.40 \times 10^{-3}$ 
& $2.26 \times 10^{-4}$  \\

$\tau_2$ 
& 68.0
& 59.2 \\

$\DOmega_p$ 
& $6.56 \times 10^{-4}$ 
& $1.21 \times 10^{-6}$ \\
\hline
\end{tabular}
\caption{Fitted parameters to equation (14) with $N = 2$ for the bulk glitches 
in Figure 8 with strong ($B = 0.1$) 
and weak ($B = 10^{-4}$) mutual friction (MF) at $Re = 500$.}
\end{table}

For both strong and weak mutual friction, the crust spins up 
immediately after recoupling.
Unlike crust glitches, where the spin up is instantaneous by construction, 
there is a noticeable rise time in $\DOmega$, which is reflected in the fitting
algorithm returning $\DOmega_1 < 0$ in Table 2.
The reason that $\DOmega$ increases initially is that spinning up the protons increases
$\partial v_p^\phi / \partial r \vert_{r = R_2}$, $N_2$ [via equation (12)],
and hence $\DOmega$ [via equation (13)].
As the crust spins up, however, the no-slip boundary condition increases 
$v_p^\phi(r = R_2)$, which acts to reduce the viscous torque and halt the spin up
after $t \approx 5$.
$\DOmega$ reaches a maximum at similar times for both strong and mutual friction
regimes ($t = 5.0$ and $t = 6.3$ respectively), which corresponds to approximately one rotation
period of the star $P = 2 \pi / \Omega_2 = 7.0$.
The spin up time is also similar to $\tMF$ for strong mutual friction, 
as discussed further in section 6.2.
For $t \gtrsim 5$, $\DOmega$ decreases on a time-scale of $\tau_2 \approx 65$.
This is because the spin up of the crust following the glitch reduces the angular 
velocity lag between the crust and core and upsets spin-down equilibrium.
In order to restore spin-down equilibrium, the crust spins down faster than the core
for $\approx 300$ time units $\sim \tvisc$, after which spin-down equilibrium is
restored.
This is similar to what we see during the set up of the initial condition.
In Figure 2, we find $\vert \dot{\Omega}_2 \vert > \vert \dot{\Omega}_1 \vert$ 
for $t \lesssim 300$, and
$\vert \dot{\Omega}_2 \vert \approx \vert \dot{\Omega}_1 \vert$ for $t \gtrsim 300$.

The evolution of $\DOmega$ is different in the strong and weak mutual
friction regimes in several ways.
Firstly, the peak size of the glitch is larger for weak than strong mutual 
friction, with 
$\DOmega_{\rm{max}}^{\rm{weak}} \approx 20 \DOmega_{\rm{max}}^{\rm{strong}}$.
Secondly, while weak mutual friction produces a fast, monotonic
rise followed by a slower, monotonic recovery, there is a noticeable
oscillatory component in both the rise and recovery for $t \leq 30 \approx \tEk$
with strong mutual friction.
The oscillation period is $\approx 5$ time units, similar to $\tMF$, which may explain
the absence of oscillation in the glitch with weak mutual friction: any oscillations
which occur on a time-scale of $\tau_{\rm MF,weak} = 500$ are invisible because 
$\DOmega(t)$ decays exponentially on a time scale of $\tau_{\rm 2, weak} = 68$
(Table 2).
Thirdly, the fitted rise time $\tau_1$ is $\sim$ 10 times faster with 
strong mutual friction than with weak, as shown in Table 2.
However, looking at Figure 8, this is an artifact of the fit;
$\DOmega$ actually peaks at $t = 5.0$ and $t = 6.3$ for strong and weak mutual 
friction respectively, which corresponds to approximately one rotation period.
Finally, weak mutual friction causes a permanent change in 
the spin frequency of the crust, viz.  
$\DOmega^{\rm weak}(t = 500) / \DOmega_{\rm max}^{\rm weak} = 0.14$,
c.f. $\DOmega^{\rm strong}(t = 500) / \DOmega_{\rm max}^{\rm strong} = 0.003$.
Though this result is not obvious in Figure 8, which truncates the recovery
at $t = 100$ for clearer presentation of the oscillation, it can be seen 
clearly in the fitted values of $\DOmega_p$ in Table 2, viz.
$\DOmega_p^{\rm weak} / \DOmega_{\rm max}^{\rm weak} = 0.14$ and
$\DOmega_p^{\rm strong} / \DOmega_{\rm max}^{\rm strong} = 0.005$.

\subsection{Inner glitches}

Inner glitches resemble crust glitches, except that $\Omega_1$ jumps initially.
The results below describe a glitch where $\Omega_1$ is instantaneously increased
to $\Omega_1(t = 0) = 1$, although in general we find that the size
of the jump does not qualitatively change the recovery dynamics.
Figure 9 plots the residual $\DOmega$ for 500 time units following the glitch.
\begin{figure}
\centering
\includegraphics[scale=0.5]{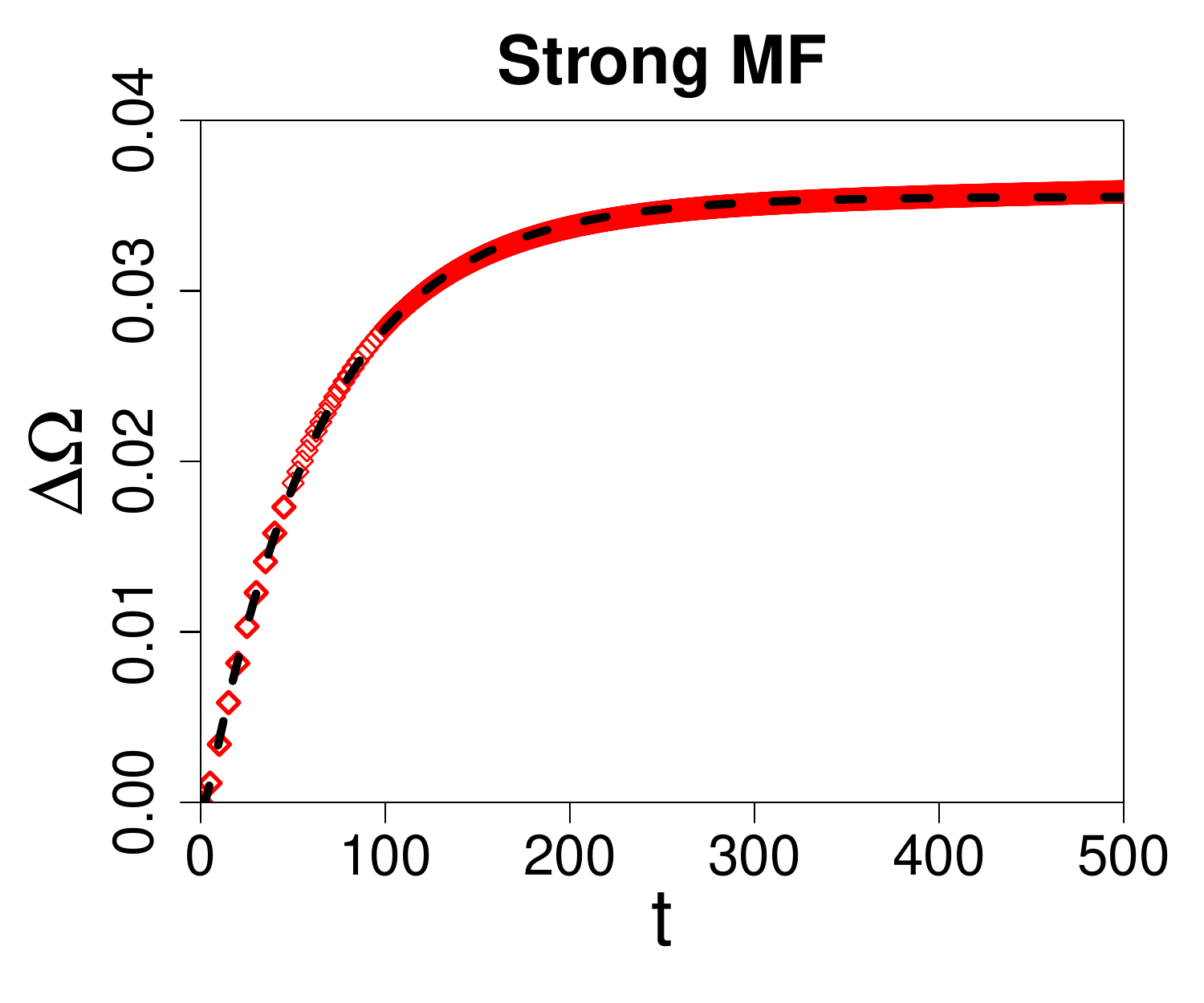} \\
\includegraphics[scale=0.5]{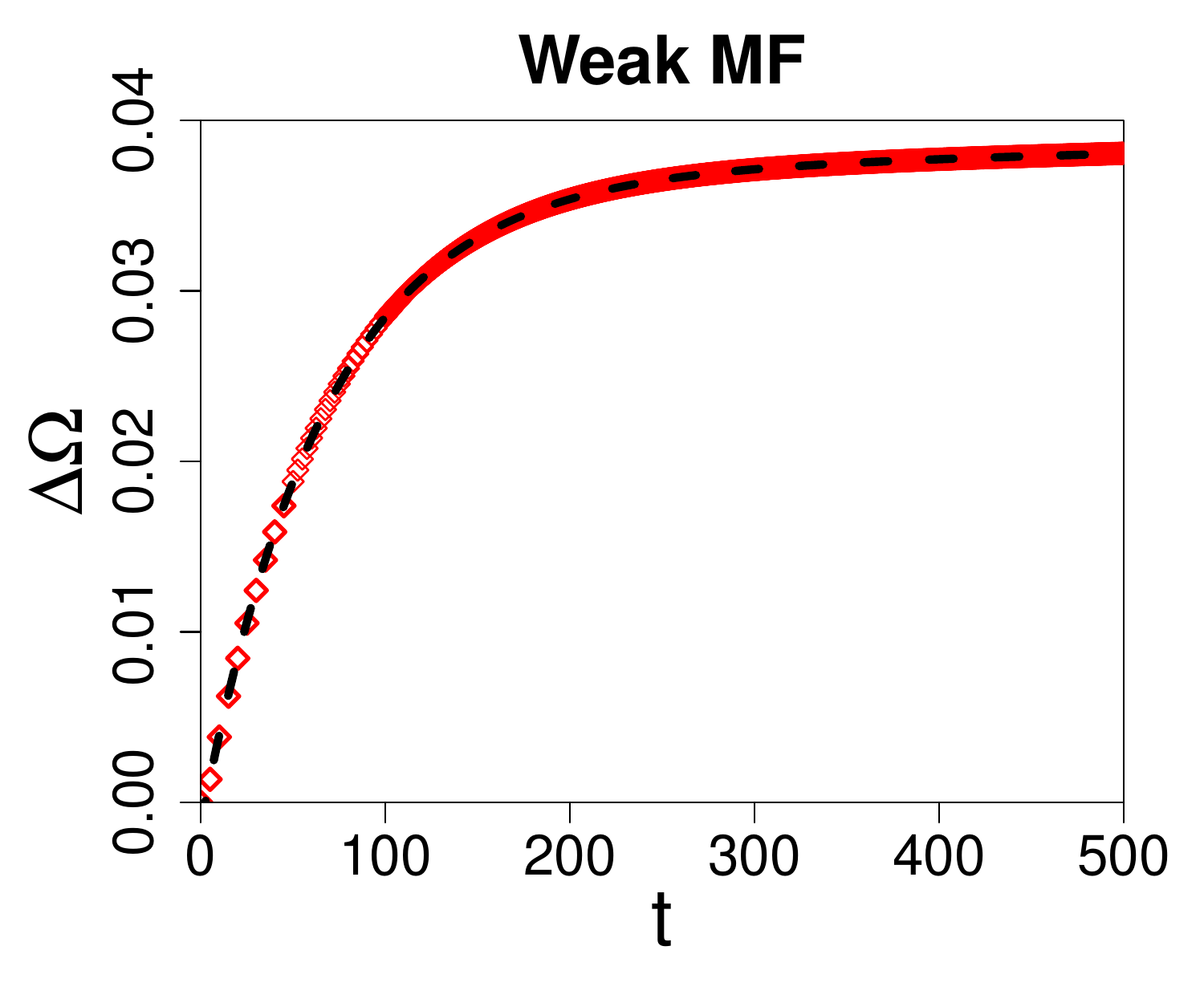}
\caption{Angular velocity residuals,
$\DOmega(t) = \Omega_{2,g}(t) - \Omega_{2,ng}(t)$, for 500 time units following
an inner glitch.
The red diamonds show the data from the simulation.
The black dashed curve shows a curve of best fit to a dual exponential recovery
given by equation (14) with $N = 2$ (see Table 3 for values).
In contrast to crust and bulk glitches, there is no relaxation following
the spin up, $\DOmega_{\rm max} = \DOmega(t = 500)$.
We see no significant difference in the recovery timescales between strong
and weak mutual friction.}
\end{figure}
The fitted parameters of a dual exponential fit are shown in Table 3.
\begin{table}
\centering
\begin{tabular}{l c c}
\hline
Parameter & Weak MF & Strong MF \\
\hline
$\DOmega_1$ 
& $1.46 \times 10^{-3}$ 
& $2.09 \times 10^{-3}$   \\

$\tau_1$ 
& 0.98 
& 1.80 \\

$\DOmega_2$ 
& $-3.90 \times 10^{-2}$ 
& $-3.73 \times 10^{-2}$  \\

$\tau_2$ 
& 70.3
& 63.5 \\

$\DOmega_p$ 
& $3.78 \times 10^{-2}$ 
& $3.55 \times 10^{-2}$ \\
\hline
\end{tabular}
\caption{Fitted parameters to equation (14) with $N = 2$ 
for the inner glitches in Figure 9 with 
strong ($B = 0.1$) and weak ($B = 10^{-4}$) mutual friction (MF) at $Re = 500$.}
\end{table}
The recovery following an inner glitch is noticeably different from crust 
and bulk glitches.
The glitch algorithm adds angular momentum to the system by increasing $\Omega_1$,
which causes $v_p^\phi(r = R_1)$ and hence $N_1$ to also increase, 
increasing the spin-down torque on the core and upsetting spin-down equilibrium. 
The system returns to spin-down equilibrium after $t \approx \tvisc$ by redistributing
the added angular momentum from the glitch between $r = R_1$ and $r = R_2$, which 
also causes the crust to spin up.
Figure 9 shows how, immediately following the glitch, $\DOmega$ rises monotonically 
and asymptotes towards a maximum value after $t \sim 300 \sim \tvisc$.
There is no relaxation.

\section{Discussion}

We discuss the above results in two groups. 
Firstly, we consider the crust and inner glitches, which can be grouped naturally,
since both involve impulsive acceleration of a boundary and a net increase of 
the angular momentum of the system (section 6.1). 
Secondly, we consider the angular momentum-conserving bulk glitches (section 6.2).

\subsection{Crust vs inner glitches}

A simple way to model the observed behaviour is by considering an 
idealised system consisting of four rigidly rotating components: 
the crust, the proton fluid, the neutron fluid, and the inner core,
all of which may couple to or decouple from one another.

To describe crust glitches in this picture, we consider the inner
core and the crust as being coupled prior to the glitch (see the discussion of
spin-down equilibrium in section 4.2 and the bottom panels of Figure 2).
The inertia of the neutrons and protons is small compared to the crust and core.
The glitch is induced by suddenly spinning up the crust, upsetting spin-down equilibrium.
A restoring force quickly spins down the crust and increases the angular velocity 
of the fluid and the core.
Eventually, spin-down equilibrium is restored, and the permanent increase
in the spin frequency of the crust can be estimated by multiplying the initial 
glitch size $\DOmega_{\rm max} = \DOmega(t = t_{\rm glitch})$ by the ratio 
of the moments of inertia of the crust and the whole system, 
$I_{\rm crust} / I_{\rm total}$.
For inner glitches the argument is the same, but the relevant ratio
is $I_{\rm core} / I_{\rm total}$, and $\DOmega_{\rm max}$ is the amount
by which the inner boundary spins up.

In Table 4 we compare the estimated permanent spin-up, 
$\DOmega^{\rm est}_p = \DOmega_{\rm max} I_{\rm crust,core} / I_{\rm total}$
to the measured value of $\DOmega$($t = 500$), as well as the fitted value 
$\DOmega_p$ for crust and inner glitches with two different spin-down models 
described below.
\begin{table*}
\centering
\begin{tabular}{l l l l l l}
\hline
Location & MF & $\DOmega$($t = 500$) 
& $\DOmega_p^{\rm est}$ & $\DOmega_p^{(5)}$ & $\DOmega_p^{(6)}$\\
& & ($\times 10^{-3}$) & ($\times 10^{-3}$) 
& ($\times 10^{-3}$) & ($\times 10^{-3}$) \\
\hline
Crust & Strong & 2.48 & 2.31 & 2.46 & 2.42 \\
Crust & Weak & 2.99 & 2.67 & 2.60 & 2.90 \\ 
Inner & Strong & 3.58 & 3.44 & 3.55 & 3.44 \\
Inner & Weak & 3.81 & 3.54 & 3.78 & 3.67 \\
\hline
\end{tabular}
\caption{We compare different methods for estimating the permanent change in the crust
angular velocity residual, $\DOmega_p$, following crust and inner glitches with 
strong and weak mutual friction.
The third column estimates $\DOmega_p$ as the value of $\DOmega$ after 500 time units of
simulation post-glitch, $\DOmega$($t = 500$).
The fourth column estimates $\DOmega_p$ using the formula
$\DOmega_p^{\rm est} = \DOmega_{\rm max} I_{\rm crust, core} / I_{\rm total}$.
Finally we estimate $\DOmega_p$ by fitting the data to glitch recovery models similar
to equation (14) with five parameters (fifth column) and six parameters (sixth column).
}
\end{table*}
For the crust glitches, Table 4 shows that the estimated values agree approximately, 
though in both mutual friction regimes we obtain $\DOmega(t = 500)$ 
$\approx 1.1 \DOmega^{\rm est}_p$.
The discrepancy arises because the system is not completely coupled, 
even after 500 time units, so that the outer boundary is decelerating faster under 
the external torque than it would in equilibrium.
In Table 4 we also consider the effect of adding a permanent change in the spin-down
rate to our glitch recovery model, as such a change is often reported in the literature
[e.g. \citet{Wang2000}].
$\DOmega_p^{(5)}$ is the fitted value implied by equation (14), while 
$\DOmega_p^{(6)}$ also includes a sixth parameter, the permanent change in 
the spin-down rate $\Delta \dot{\Omega}_p t$ in equation (14).
Both values agree to within 10\%, both in comparison to each other and also to the 
predicted and observed values, and the sum-of-squares errors returned by the 
fitting algorithm for each model are similar. 

\subsection{Bulk glitches and mutual friction}

An interesting result from the bulk glitch simulations is the large size 
difference between glitches with strong and weak mutual friction.
Figure 5 shows that $\absvpnphi$ is greater for weak 
mutual friction than strong, but the difference is at most a factor
of $\approx 4$, not the factor of $\approx 20$ observed in glitch sizes.
To understand this discrepancy, we treat both fluid components and the boundaries 
as rigid bodies, so that the angular momentum of each is given by $I_x \Omega_x$,
where the index $x =$ p, n or b denotes denote the protons, neutrons and 
boundary respectively.
Conservation of angular momentum before and after the glitch implies
\begin{equation}
I_p \Omega_p^i + I_n \Omega_n^i + I_b \Omega_b^i =
I_p \Omega_p^f + I_n \Omega_n^f + I_b \Omega_b^f \, ,
\end{equation}
where the superscripts $i$ and $f$ denote the initial (pre-glitch) 
and final (post-glitch) values respectively.
Rearranging (17), we get
\begin{equation}
\DOmega_b = -\frac{I_p}{I_b} \DOmega_p - \frac{I_p}{I_b} \DOmega_n \, ,
\end{equation}
where $\DOmega_x = \Omega_x^f - \Omega_x^i$.
If the proton fluid couples to the boundary much faster than the neutrons,
so that the boundary spins up before being the neutrons recouple, 
then the effect of the neutrons can be neglected, and we can estimate the maximum
glitch size as 
\begin{equation}
\DOmega_{\rm max}^{\rm est} = \frac{\rho_p I_3\vpnphiav^{\rm max}}{2 I_2}  \, .
\end{equation}
For weak mutual friction, equation (19) yields 
$\DOmega_{\rm max}^{\rm est} = 3.8 \times 10^{-3}$, compared to
$\DOmega_{\rm max} = 4.6 \times 10^{-3}$ from the simulation.
For strong mutual friction, the prediction is worse, 
$\DOmega_{\rm max}^{\rm est} = 1.2 \times 10^{-3}$, versus
$\DOmega_{\rm max} = 2.3 \times 10^{-4}$ from simulation.
That the glitch size is overestimated with strong mutual friction is 
unsurprising, since setting $\DOmega_n = 0$ in equation (19) implicitly assumes
that the spin-up time is much greater than $\tMF$, whereas actually the spin-up time 
is $\approx$ one rotation period $\approx \tMF^{\rm strong}$ [section 5.4].
In this regime, mutual friction suppresses spin up, limiting the size of a glitch.
The oscillations with period $\approx \tMF^{\rm strong}$ seen in the top panel of
figure 8 may be related to the reduction in glitch size with strong mutual friction.

Another interesting result from the bulk glitch simulations is that weak 
mutual friction produces a permanent change in $\DOmega$, while strong mutual 
friction restores the pre-glitch trend after 500 time units.
During spin down, the neutron condensate is not affected by the spin-down torque
initially, and only spins down with the crust and viscous component after a 
time $\sim \tMF$.
At $t \lesssim \tMF$, the condensate spins down slower than the crust and 
viscous components, building up a `reservoir' of angular momentum,
$J_{\rm res} = I_n \absvpnphi/R_2$.
With weak mutual friction, this reservoir grows at all times during the 
simulation, while with strong mutual friction it saturates.

When the glitch occurs, angular momentum is transferred from the `reservoir' 
to the protons, which then spin up the crust. 
As discussed in section 5.3, spinning up the crust also increases the spin-down rate,
so $\DOmega$ decreases for $t \gtrsim 5$, until spin-down equilibrium is restored
after $t \approx \tvisc = 500$.
Following a bulk glitch, the angular momentum of the overall system is unchanged, 
but the `observable' $\DOmega$ is the crust frequency, and the crust only couples to 
the neutrons when mutual friction is strong.
When mutual friction is weak, the neutrons are decoupled from the crust, so the 
decrease in $v_n^\phi$ accompanying the increase in $v_p^\phi$ cannot be seen in 
$\DOmega$, and the result of the glitch is an increase in the angular momentum 
of the observable components in the system.

This result has interesting implications in the context of the healing parameter,
usually defined as 
\citep{Wang2000}
\begin{equation}
Q = 1 - \frac{\DOmega_p}{\DOmega_{\rm max}} \, ,
\end{equation}
so that $Q \rightarrow 0$ for a glitch that recovers fully, and $Q \rightarrow 1$
for a glitch with no recovery.
Taking $\DOmega_{\rm p} = \DOmega(t = 500)$, we find $Q = 0.143$ with weak
mutual friction and $Q = 0.003$ with strong mutual friction.
Generally, $Q$ is thought to be a measure of the relative moments of inertia of
the inviscid and viscous components locked to the crust, but our results above
indicate that mutual friction is important.
Specifically, our results affect the standard argument, which holds that if
some fraction of the superfluid with moment of inertia $I_{\rm res}$ spins down
slower than the crust and viscous components, whose total moment of inertia is denoted 
$I_c$, then the ratio $I_{\rm res} / I_c$ provides an upper bound on the angular 
momentum which is `recovered' by glitches, i.e. if $\dot{\Omega}_{\rm sd}$ is the
average spin-down rate of a pulsar and $\dot{\Omega}_{\rm glitch}$ is the average
spin-up rate from glitches then 
$I_{\rm res} / I_c \geq \dot{\Omega}_{\rm glitch} / \dot{\Omega}_{\rm sd}$.
Often [e.g. \citet{Link1999, Andersson2012}],
$I_{\rm res}$ estimated in this way is taken to be an indication of the superfluid fraction in the inner crust, is then used to put constraints on the
nuclear equation of state. 
However, our simulations show that mutual friction affects the size of the 
reservoir more than $\rho_n / \rho_p$.

\section{Conclusions}

In this paper, we perform numerical simulations of pulsar glitches 
by solving the two-fluid equations of motion for the neutrons and protons
in spherical Couette geometry.
Glitches are simulated in three ways: firstly, by an instantaneous increase
in the angular velocity of the crust; secondly, by an instantaneous increase
in the angular velocity of the inner core; and thirdly, by instantaneously
locking together the neutron and proton velocities throughout the interior.
All three experiments start from a realistic initial state which is set up 
from initially corotating boundaries, after which the outer boundary is spun down 
by a constant torque on the crust for $\approx 10^2$ rotation periods.
In all cases we observe that the angular velocity of the crust increases,
but we find that the response of the crust angular velocity following a glitch
varies depending on the way the glitch is activated.
Glitches that originate in the crust exhibit an instantaneous angular velocity jump,
followed by an exponential relaxation towards the pre-glitch trend.
Glitches that originate in the core exhibit a permanent angular velocity increase
building up over $\approx 50$ rotation periods with no subsequent relaxation.
Glitches that are caused by a sudden recoupling of the two fluid components
in the bulk have a rise time that is similar to the rotation period of the star, 
followed by an exponential relaxation.
Glitch sizes and the smoothness and completeness of the recovery are affected by 
the strength of mutual friction.
The finding that different glitch activation mechanisms produce different kinds of 
recoveries may help to explain some of the diversity seen in the population 
of glitches 
\citep{Wong2001, Espinoza2011, Haskell2014}.

Our results demonstrate the importance of mutual friction in bulk glitches.
When mutual friction is strong, the lag between the two fluid components can 
reverse between $r = R_1$ and $r = R_2$, so that the neutron condensate spins 
down faster than the protons in some parts of the star.
Mutual friction also affects the size of bulk glitches: with stronger mutual friction
glitches are smaller, and the recovery is non-monotonic, exhibiting oscillations
with a period similar to the mutual friction time-scale which are damped after an Ekman
time. 
As well as these interesting effects, we also find that mutual friction 
controls the healing parameter $Q$, a result which has implications for using 
pulsar glitches to test models of nuclear matter
\citep{Link1999, Andersson2012, Newton2015}.

In order to preserve numerical stability and work with available computationalresources,
we necessarily make a number of simplifying assumptions.
In particular, the choice of superfluid boundary conditions is an open question in the
literature, which will be refined in light of future experimental and theoretical work.
We also aim to improve the model in the future by including entrainment and 
vortex tension in the equations of motion and adapting the two-fluid solver to 
allow for non-uniform densities.

\section*{Acknowledgments}

This work was supported by Australian Research Council (ARC) Discovery Project Grant 
DP110103347.
G. H. would like to thank Dr Carlos Peralta and Dr Eric Poon for their assistance
with numerical simulations.
G. H. acknowledges support from the University of Melbourne through via 
Melbourne Research Scholarship.
B. H. acknowledges support from the ARC via a Discovery Early Career Researcher Award.
Simulations were performed on the MASSIVE cluster\footnote{https://www.massive.org.au/}
at Monash University, using computer time awarded through the 
National Computational Merit Allocation Scheme.

\bibliographystyle{mn2e}
\bibliography{glitchrecovery}

\label{lastpage}

\end{document}